\def\calb  {{\cal B}}
\def\calc  {{\cal C}}
\def\cald  {{\cal D}}
\def\cale  {{\cal E}}
\def\calf  {{\cal F}}
\def\calg  {{\cal G}}
\def\calh  {{\cal H}}

\def\calk  {{\cal K}}

\def\calm  {{\cal M}}
\def\caln  {{\cal N}}

\def\calr  {{\cal R}}
\def\cals  {{\cal S}}
\def\calt  {{\cal T}}
\def\calu  {{\cal U}}
\def\calv  {{\cal V}}

\def\caly  {{\cal Y}}

\def\dl            {\mathbb }
\def\complex       {{\dl C}}

\def\rationals     {{\dl Q}}
\def\reals         {{\dl R}}
\def\zet           {{\dl Z}}

\newcommand\hsp[1] {\mbox{\hspace{#1 em}}}
\newcommand\nxt[1] {\\\raisebox{.12em}{\rule{.35em}{.35em}}\hsp{.6}#1}
\long\def\query#1{\hskip 0pt{\vadjust{\everypar={}\small\vtop to 0pt{\hbox{}%
     \vskip -13pt\rlap{\hbox to 49.0pc{\hfil{\vtop{\hsize=8pc\tolerance=6000%
     \hfuzz=.5pc\rightskip=0pt plus 3em\noindent#1}}}}\vss}}}}%
\def\alg           {algebra}
\def\auto          {automorphism}
\newcommand\Barray[2]{\mbox{\large$[$}\!{\scriptstyle\begin{array}{c}{}\\[-1.94em]
                   {\scriptstyle #1}\\[-.43em]{\scriptstyle #2}\\[-.4em] \end{array}}\!%
                   \mbox{\large$]$}}

\def\be            {\begin{equation}}
\def\bearl         {\begin{array}{l}}
\def\bearll        {\begin{array}{ll}}
\def\bfe           {{\bf1}}
\def\bigtimes      {\mbox{\LARGE$\times$}\!}
\def\cft           {conformal field theory}

\def\chii          {\raisebox{.15em}{$\chi$}}
\def\ee            {\end{equation}}
\def\eE            {{\rm e}}
\def\eear          {\end{array}}
\def\eps           {\epsilon}
\def\eq            {\,{=}\,}
\newcommand\erf[1] {(\ref{#1})}
\newcommand\Erf[2] {(\ref{#1#2})}
\newcommand\F[6]   {{\sf F}_{\! #1,#2}\Barray{#4\,#5}{#3\,#6}}
\def\findim        {fi\-ni\-te-di\-men\-si\-o\-nal}
\newcommand\Frac[2]{\mbox{\large$\frac{#1}{#2}$}}
\def\g             {{{\mathfrak g}}}
\def\Homeo         {{\rm Homeo}}
\def\hy            {$\mbox{-\hspace{-.66 mm}-}$}
\def\id            {{\rm id}}
\def\ii            {{\rm i}}
\def\iN            {\,{\in}\,}
\def\infdim        {in\-fi\-ni\-te-di\-men\-si\-o\-nal}

\def\irrep         {irreducible representation}
\def\jv            {\mbox{$J_{\rm v}$}}
\def\js            {\mbox{$J_{\rm s}$}}

\long\def\labl#1   {\label{#1}\ee \ifnum\draftcontrol=1
                   \mbox{ }\\[-12 mm]\query{#1}\\[5 mm] \fi}
\long\def\Labl#1#2 {\label{#1#2}\ee\ifnum\draftcontrol=1
                   \mbox{ }\\[-12 mm]\query{#1#2}\\[5 mm] \fi}
\def\lie           {Lie algebra}
\newcommand\N[3]   {{\cal N}_{#1 #2}^{\;\ #3}}
\def\onedim        {one-dimen\-sional}
\def\ot            {\raisebox{.07em}{$\scriptstyle\otimes$}}
\def\oT            {\,\ot\,}
\def\ots           {\raisebox{.07em}{$\sss\otimes$}}
\newcommand\pso[4] {\Psi_{\;#1}^{#2\,{\sss{\rm #4}}\,#3}}

\def\rep           {representation}
\def\rmd           {{\rm d}}
\def\sicu          {\mbox{$J$}}
\def\sss           {\scriptscriptstyle }
\newcommand\suma[4]{\sum_{{\rm #1}=1}^{\N{#3}{#4}{#2}}}
\newcommand\sumn[4]{\sum_{{\rm #1}=1}^{\N{#2}{#3}{#4}}}
\newcommand\sxs[3] {{{\sss{\rm #1}}}{#2}{{\sss{\rm #3}}}}
\newcommand\tC[9]  {C_{\,#1\,{\sss{\rm #6}}\,#2}^{#3\,{\sss{\rm #7}}
                   \,#4\,{\sss{\rm #8}}\,#5\,{\sss{\rm #9}}}}

\def\twodim        {two-di\-men\-si\-o\-nal}
\def\twtw          {^{\sss(2)}}
\def\untw          {^{\sss(1)}}
\def\Vir           {{\rm Vir}}

\def\wzwts         {WZW theories}
\catcode`\@=11
                             \global\def\prlcontrol{1}
                             \global\def\prlcontrol{0}
 \def\incomplete{\ifnum\prlcontrol=0
     \query{\bf to be\\\mbox{}$\!\!\!\!$completed} \fi}

\newif\if@fewtab\@fewtabtrue
{\count255=\time\divide\count255 by 60
\xdef\hourmin{\number\count255}
\multiply\count255 by-60\advance\count255 by\time
\xdef\hourmin{\hourmin:\ifnum\count255<10 0\fi\the\count255}}
\def\ps@draft{\let\@mkboth\@gobbletwo
    \def\@oddhead{}
    \def\@oddfoot{\hbox to 7 cm{\tiny \versionno
       \hfil}\hskip -7cm\hfil\rm\thepage \hfil {\tiny\draftdate}}
    \def\@evenhead{}\let\@evenfoot\@oddfoot}
\def\draftdate{\number\month/\number\day/\number\year\ \ \ \hourmin }

\global\def\draftcontrol{0}

\def\draftcite#1{\ifnum\draftcontrol=1#1\else{}\fi}
\def\@lbibitem[#1]#2{\item{}\hskip -3\hbox to 2cm
{\hfil$\scriptstyle\draftcite{#2}$}\hskip
1cm[\@biblabel{#1}]\if@filesw
     {\def\protect##1{\string ##1\space}\immediate
      \write\@auxout{\string\bibcite{#2}{#1}}}\fi\ignorespaces}

\def\@bibitem#1{\item\hskip -3cm \hbox to 2cm
{\hfil {\footnotesize\draftcite{#1}}}\hskip 1cm
\if@filesw \immediate\write\@auxout
       {\string\bibcite{#1}{\the\value{\@listctr}}}\fi\ignorespaces}

\documentclass[12pt]{article} \usepackage{amssymb,amsfonts,latexsym}
\usepackage[dvips]{graphics} \usepackage{epsfig}
\begin{document}

\begin{flushright}  {~} \\[-1cm]
{\sf hep-th/0011109}\\{\sf PAR-LPTHE 00-41}\\{\sf CERN-TH/2000-328}\\
{\sf ETH-TH/00-10} \\ {\sf IHP-2000/02}
\\[1mm]
{\sf November 2000} \end{flushright}

\begin{center} \vskip 14mm
{\Large\bf CONFORMAL FIELD THEORY,} \\[4mm]
{\Large\bf BOUNDARY CONDITIONS AND} \\[4mm]
{\Large\bf APPLICATIONS TO STRING THEORY}\\[20mm]
{\large Christoph Schweigert$\;^1$\, J\"urgen Fuchs$\;^2$ \ and \ 
Johannes Walcher$\;^{3,4}$}
\\[8mm]
$^1\;$ LPTHE, Universit\'e Paris VI~~~{}\\
4 place Jussieu\\ F\,--\,75\,252\, Paris\, Cedex 05\\[5mm]
$^2\;$ Institutionen f\"or fysik~~~~{}\\
Universitetsgatan 1\\ S\,--\,651\,88\, Karlstad\\[5mm]
$^3\;$ Theory Division, CERN\\ CH\,--\,1211 Gen\`eve 23 \\[5mm]
$^4\;$ Institut f\"ur Theoretische Physik, ETH H\"onggerberg,\\ 
CH\,--\,8093 Z\"urich 
\end{center}
\vskip 18mm
\begin{quote}{\bf Abstract}\\[1mm]
This is an introduction to two-dimensional conformal field theory and its 
applications in string theory. Modern concepts of conformal field theory are 
explained, and it is outlined how they are used in recent studies of D-branes 
in the strong curvature regime by means of CFT on surfaces with boundary.
\end{quote}
\newpage

\section{Introduction and Overview}

Conformal field theory in two dimensions (CFT) is an old subject. It is a rich 
subject, too. Several points of view  on CFT are possible. In the present 
lecture notes, we adopt one perspective that is suitable for the application 
of conformal field theory to (perturbative) string theory. Thus in particular
we will always work with compact world sheets of Euclidean signature.
(The signature of the world sheet should not be confused with the signature
of space-time; a consistent string theory even requires time-like directions 
in the target space.)

To understand CFT it is necessary to clearly distinguish between {\em chiral\/}
and {\em full\/} \cft. This distinction is, unfortunately, not always respected
in the literature. Later in these lectures we will explain in some detail what 
we understand by chiral and full CFT. For the moment, the reader only needs to 
retain the following features. The two theories are defined on different types 
of \twodim\ manifolds: 
\nxt{}{\em Chiral\/\em} CFT is defined on complex curves. It is the theory
     that appears in the description of the (fractional) quantum Hall effect.
     And it is also chiral CFT that is related to topological field theory and
     that, as a consequence, provides invariants of knots and three-manifolds.
\nxt{}{\em Full\/\em} CFT, in contrast, is defined on real \twodim\ manifolds
     with a conformal structure, i.e.\ an equivalence class, with respect to 
     local rescalings, of metrics. The manifold can have boundaries, and need 
     not be orientable. Notice that even when it is orientable, no definite
     orientation is chosen. 

Full CFT has numerous applications. The point that is of prime interest here
is that it provides the world sheet
theories for string theory. Other applications concern 
two-dimen\-si\-o\-nal critical systems of classical statistical mechanics
\cite{Card} and quasi one-dimensional condensed matter physics \cite{Affl}.
Also, CFT models that are based on a Lagrangian are generally models of full 
CFT. Note that for writing down a Lagrangian, the world sheet must typically 
have a Hodge star which is needed for the conformal structure,
but it is not necessary to have a metric on the world sheet. Frequently the
Lagrangian formulation is in the form of a sigma model, i.e.\ the
path integral is taken over maps $X$ from the world sheet to some manifold, 
the {\em target space\/}. (See e.g.\ \cite{Gawe} for a review of 
Wess\hy Zumino\hy Witten models from this point of view.) 
{}From any chiral CFT one can construct a full CFT. Conversely, it can
be expected that {\em every\/} consistent conformal field theory in two 
dimensions comes from an underlying chiral CFT; in all applications of CFT we 
are aware of, such a relationship is known in detail.

We will be particularly interested in the study of world sheets with boundaries.
Open string perturbation theory in the background of certain string solitons, 
the so-called D-branes, indeed forces one to
analyze conformal field theories on surfaces that may have boundaries
and\,/\,or can be non-orientable. Other applications of CFT on surfaces with
boundary -- which are, however, beyond the scope of these notes -- arise in
the study of defects in systems of condensed matter physics, of percolation 
probabilities, and in dissipative quantum mechanics.

These notes are organized as follows. We first review in Section 2 aspects 
of two-dimensional manifolds that will enter our discussion. We then turn to 
the analysis
of chiral conformal field theory. After discussing the underlying algebraic
structure -- vertex operator algebras (VOAs, for short) -- in Section 3, we 
proceed in Section 4 to combine VOAs and the theory of complex curves,
which enables us to define conformal blocks.
The latter are the central objects in a chiral CFT. Their physical role
is twofold. On one hand, they are building blocks for the correlators 
of full CFT, and on the other hand they are the spaces of physical states
in topological field theories (TFTs) in three dimensions.

After this discussion of chiral CFT, in Section 5 we explain the construction 
of a full CFT from a chiral CFT. In Section 6 it is shown how
to construct (perturbative) string vacua from full CFTs with appropriate
properties. The consideration of full CFT on world sheets with
boundaries will also give us some insight into what is sometimes called
`D-geometry' \cite{doug8}, that is, the physics of strings and branes for 
finite values of the string length, i.e.\ in the region
where curvature effects become important.

We conclude this introductory section with a word on the citation policy
adopted in these lecture notes. We decided to mention mainly review 
papers or papers that have been helpful for ourselves to improve our 
understanding of the topics treated here and of their physical and 
mathematical background. Earlier references are only included when they 
treat issues for which we do not know of an accessible review; otherwise we 
refer the reader to those reviews as a guide to original work. 
(A more extensive bibliography on CFT can be found at {\tt 
http:/$\!$/home.cern.ch/$\tilde{\phantom{x}}$jfuchs/ref.html}; for
lack of space we have refrained from including most of those references here.)

\section{Two-dimensional manifolds}

Manifolds of (real) dimension two constitute the `arena' for our studies.
In this section we briefly summarize those features that are needed in our 
discussion. We start with topological aspects, then discuss complex geometry 
of these manifolds, and finally describe Teichm\"uller and moduli spaces.

\subsection{Topological aspects}

Real connected compact two-dimensional topological manifolds $\Sigma$ are
classified by three non-negative integers: The numbers $g$ of handles,
$b$ of boundaries, and $c$ of crosscaps. 

By cutting out a disc from a given surface one introduces a new
boundary component.  (We insist that the boundaries obtained this way are 
{\em genuine, physical boundaries\/}. 
They must never be confused with the small discs one often
imagines around insertions of fields, which merely serve to specify
local coordinates around the insertion points.) Similarly, a crosscap can be
inserted in a surface by first cutting out a disc and then identifying
opposite points of the boundary of the disc. The insertion of a crosscap
makes a manifold unorientable. Three crosscaps
are topologically equivalent to a single crosscap plus a handle, so that
it is not necessary to consider more than two crosscaps. 
World sheets with boundaries play an important role in the
description of D-branes; world sheets with crosscaps enter in the 
construction of string theories of `type I'.

An important topological quantity is the Euler characteristic $\chi$.
It is defined as $\chi\eq2\,{-}\,2g\,{-}\,b\,{-}\,c\,$. Its relevance 
in string perturbation
theory stems from the fact that the contribution of a surface of
Euler characteristic $\chi$ to the perturbation expansion is weighted with a 
factor ${\rm g}_{\rm str}^{-\chi}$, where ${\rm g}_{\rm str}$ is the string 
coupling constant. The most important manifolds in our discussion will 
be the ones of Euler characteristic $\chi\eq2$, i.e.\ the two-sphere
$S^2$, of $\chi\eq1$, i.e.\ the disc and the real projective space 
$\reals {\dl P}^2$ (also called the crosscap), and the manifolds with 
$\chi\eq0$, i.e.\ the torus $(g\eq1,\,b\eq c\eq0)$, the annulus $(g\eq c\eq0,\,
b\eq2)$, the M\"obius strip $(g\eq0,\,c\eq b\eq1)$, and the Klein bottle 
$(g\eq b\eq 0,\,c\eq2)$.

Another notion we will frequently use is the one of the {\em mapping class 
group\/} $\Omega$ of $\Sigma$. 
Consider the group $\Homeo(\Sigma)$ of all homeomorphisms of $\Sigma$.
It has a normal subgroup $\Homeo_0(\Sigma)$, consisting of those homeomorphisms 
that are homotopic to the identity. If $\Sigma$ is orientable, we introduce the 
subgroup $\Homeo^+(\Sigma)$ of orientation preserving homeomorphisms.
For orientable surfaces, we define the mapping class group as the quotient
  $$ \Omega(\Sigma) := \Homeo^+(\Sigma) / \Homeo_0(\Sigma) \,, $$
while for unorientable surfaces we define it to be
  $$ \Omega(\Sigma) := \Homeo(\Sigma) / \Homeo_0(\Sigma) \,. $$

The mapping class group $\Omega(\Sigma)$ acts in particular on the
first homology $H_1(\Sigma,\zet)$. For orientable surfaces without boundary 
$H_1(\Sigma,\zet)$ is torsion free and comes with a symplectic form from
the intersection of one-cycles. (For unorientable surfaces $H_1(\Sigma,\zet)$ 
typically has a torsion part.) The mapping class group preserves the
intersection form, and hence we obtain a natural group homomorphism 
from $\Omega(\Sigma)$ to the corresponding symplectic
group; this homomorphism is actually surjective
so that we have
  $$ \Omega(\Sigma) \to\!\!\!\!\!\to {\rm Sp}(2g,\zet) \, . $$
For the torus, we even have identity:
  $$ \Omega(T^2) = {\rm Sp}(2,\zet) = {\rm SL}(2,\zet) \, . $$
This group is called the {\em modular group}.

\subsection{The Schottky double}

We now turn to a construction that allows us to restrict our attention
to the case when the manifold $\Sigma$ is oriented and has no boundary: 
The Schottky double $\hat\Sigma$ of $\Sigma$. The idea is to 
double the space, except for the points on the boundary.
(This mimics the method of mirror charges in classical electrodynamics.)

For $\Sigma$ a disc, the double $\hat\Sigma$ is a sphere, obtained by 
gluing a disc and its mirror image along their boundaries. Notice that the 
reflection $\sigma$ about the equator of this sphere is an 
orientation-reversing involution, $\sigma^2\eq\id$.
The original surface $\Sigma$ can be obtained as the quotient -- or world 
sheet orbifold, or parameter space orbifold -- of the double:
  \be  \Sigma = \hat\Sigma / \sigma \, . \ee
The fixed point set of $\sigma$ gives precisely the boundary of $\Sigma$.  
For $\Sigma$ the crosscap $\reals{\dl P}^2$,
the double is again the sphere, but $\sigma$ is now the antipodal map. 

It is in fact true in general that $\Sigma$ is obtained from its cover
$\hat \Sigma$ as the fixed point set under an orientation-reversing 
involution. Furthermore, $\hat \Sigma$ 
has a natural orientation. For example, for $\Sigma$ without boundary, 
the double is just the total space of the orientation bundle. (The orientation 
bundle is a $\zet_2$-bundle over $\Sigma$ whose fiber over $p\iN\Sigma$ 
consists of two points, corresponding to the two local orientations 
at $p$. Thus for orientable boundaryless $\Sigma$ it is a trivial bundle, 
the total space being the disconnected sum of two copies of $\Sigma$. 
In the results about complex curves $\hat X$ quoted below it will often be
assumed that the curve is connected; to pertain to a general CFT situation, 
those results must be applied for each of the components of $\hat X$.)
The total space of the orientation bundle is not only orient{\em able\/}, 
but even naturally orient{\em ed\/}. The Euler characteristic $\hat\chi$ of 
the Schottky cover $\hat\Sigma$ is related to the Euler characteristic 
$\chi\eq2\,{-}\,2g\,{-}\,b\,{-}\,c$ of $\Sigma$ by $\hat\chi\eq2\chi$.

\smallskip
The structures we have met so far all belong to the realm of
two-di\-men\-sional topological manifolds. We now turn to
aspects of conformal and complex manifolds that we will need
for CFT. Thus let $X$ be a two-dimensional conformal manifold,
and $\hat X$ its (topological) Schottky double.
The following idea turns out to be central:
\begin{quote}
{\sl Full CFT on a conformal manifold $X$ is constructed\\ from 
chiral CFT on the double $\hat X$ of $X$.}
\end{quote}
To understand this construction, we will also need tools from complex geometry;
they are the subject of the next subsection.

\subsection{Complex geometry}

We will be particularly interested in two-dimensional manifolds that are
even complex manifolds, i.e.\ that possess a holomorphic structure.
(A holomorphic structure is the choice of an atlas with maps that take their
values in subsets of the complex plane in such a manner that the transition 
functions are holomorphic.) On such manifolds, all the additional
power of complex geometry is at our disposal.

The following feature is special to two dimensions: A holomorphic structure on 
an orientable manifold $X$ is equivalent to the choice of a conformal
structure plus an orientation. 
This can be seen as follows. A classical theorem asserts that every metric on 
a real two-dimensional differentiable manifold is locally conformally flat,
i.e.\ we can find charts $U$ such that the metric is of the form $g\eq\lambda
(x,y)\, (\rmd x^2{+}\,\rmd y^2)$ with $\lambda$ a positive real function
on $U$. When $X$ is oriented, we can choose the charts to be compatible with
the orientation. The transformations between different charts are then
oriented and conformal diffeomorphisms, i.e.\ biholomorphic transformations,
so we have obtained a complex structure on $X$; this complex structure depends 
only on the conformal equivalence class of the metric. Conversely, it is
easy to see that a complex structure implies an orientation and a conformal
structure. In contrast, higher-dimensional manifolds are not necessarily
locally conformally flat. Accordingly, additional integrability conditions must 
then be met to obtain a complex structure.

Now the double $\hat X$ of a conformal manifold $X$ is oriented and 
inherits a conformal structure from $X$. In other words, $\hat X$ always has 
a complex structure, i.e.\ the double is a complex curve! This simple 
observation allows us to construct full CFT on $X$ in terms of chiral CFT on 
$\hat X$. In particular, at the level of chiral CFT the full power of 
holomorphy is available, even for the study of open strings. 

Once we have a complex structure on a two-manifold $X$, we can introduce
the holomorphic cotangent bundle $\calk$, also called the {\em canonical 
bundle\/} of $X$. $\calk$ is a complex line bundle, and its sections are
proportional to $\rmd z$; its powers $\calk^{\otimes n}$ are
complex line bundles as well.

\subsection{Teichm\"uller space and moduli space}

One and the same oriented topological manifold $\hat\Sigma$ without boundary
can typically be 
endowed in different, inequivalent ways with a complex structure. In fact, 
a lot of important information about CFTs is gained by studying how structures
change when one varies the underlying curve. Points at which the underlying 
curve degenerates in not too bad a way are of special importance; they lead to
{\em factorization constraints\/}.

We first consider the space $\tilde\calm(\hat\Sigma)$ of all complex structures 
$\calc$ on an orientable manifold $\hat\Sigma$. On the space 
$\tilde\calm(\hat\Sigma)$ the group $\Homeo(\hat\Sigma)$ acts as follows. The 
complex structure $f^*(\calc)$ is defined by the requirement that the map
  $$ f:\quad (\hat\Sigma,f^*(\calc)) \to (\hat\Sigma,\calc) $$
is holomorphic if $f$ preserves the orientation and antiholomorphic if
$f$ reverses the orientation. One then defines the {\em Teichm\"uller space\/}
$\calt(\hat\Sigma)$ as the quotient
  $$ \calt(\hat\Sigma):= \tilde\calm(\hat\Sigma) / \Homeo_0(\hat\Sigma) \, . $$
One can show that the Teichm\"uller space is a complex manifold. 
For $\hat g\eq0$
it is just a point, while for $\hat g\eq 1$ it is isomorphic to the complex 
upper half plane, $\calt_1\eq H\,{:=}\, \{\tau\iN\complex \,|\, {\rm Im}\tau
\,{>}\,0\}$. For every $\hat g\,{\geq}\,2$ the Teichm\"uller space is 
isomorphic to $\complex^{\,3\hat g-3}$. 
(To be precise, it is isomorphic to $\complex^{\,3\hat g-3}$ as a topological
manifold, and to $\reals^{6\hat g-6}$ as a real analytic manifold, but not
to $\complex^{\,3\hat g-3}$ as a complex analytic manifold.)

On the Teichm\"uller
space, the mapping class group $\Omega(\hat\Sigma)$ acts as the full group
of holomorphic automorphisms of $\calt(\hat\Sigma)$. The {\em moduli space\/} 
$\calm_{\hat g}(\hat\Sigma)$ is obtained from the Teichm\"uller space
by dividing out the action of $\Omega(\hat\Sigma)$:
  $$ \calm_{\hat g}(\hat\Sigma) = \calt(\hat\Sigma)/\Omega(\hat\Sigma)
  =\tilde\calm(\hat\Sigma)/\Homeo^+(\hat\Sigma) \,. $$            
If $\sigma$ is an orientation reversing homeomorphism of $\hat\Sigma$,
it induces by the same procedure an {\em anti-holomorphic\/} involution
$\sigma^*$ on the Teichm\"uller space. This will be important in the
discussion of the double. One can show that for $\hat\Sigma$ of genus
$\hat g$, there are $[\frac{3\hat g+4}2]$ inequivalent involutions.

The Teichm\"uller space is simply connected and is in fact the 
universal covering space of moduli space. This implies that
the mapping class group is the fundamental group of the moduli space:
  \be \pi_1(\calm_{\hat g}) = \Omega(\hat\Sigma) \,.  \labl{1.4}
(As one is dealing with singular spaces, one must be careful with the
definition of the fundamental group. For details, see \cite{hain}.)
For $\hat g\eq1$, the action of $\Omega$ is the standard action
  \be  \tau \;\mapsto\; \left(\begin{array}{ll}a&b\\[-.13em]c&d\end{array}
  \right) \tau \equiv \frac{a\tau+b}{c\tau+d} \ee
of SL$(2,\zet)$ on the upper half-plane $H$. The action of the mapping class 
group is not free, and as a consequence the moduli space $\calm_{\hat g}$
has {\em orbifold\/} singularities, which correspond to curves with non-trivial
automorphisms. For genus one, these are the points $\tau\eq\ii$ and
$\tau\eq\exp(2\pi\ii/3)$. However, these singular points are not the most
interesting ones for \cft.\,%
 \footnote{~See, however, a recent application of curves with automorphisms
 to derive constraints on the representation of the mapping class group
 on conformal blocks \cite{bant13}.}
Of far more interest are singularities that are cusps like the point 
$\tau\eq\ii\infty$ for $\hat g\eq1$, where the curve degenerates and where
factorization constraints can be formulated.

\medskip

We also need Teichm\"uller spaces for two-dimensional manifolds $\Sigma$
that are not oriented. $\Sigma$ can even be unorientable and 
have a boundary. Without an orientation, the notion `holomorphic' does not
make sense, so we replace in the definition of the Teichm\"uller space
$\calm(\Sigma)$ by the space of all {\em di-analytic} structures.
(A di-analytic structure is an atlas in which the transition functions
are either holomorphic or anti-holomorphic.) Teichm\"uller and moduli
space are then defined analogously as before.

Consider now the double $\hat \Sigma$ of $\Sigma$, together with the 
orientation reversing involution $\sigma$ such that $\hat \Sigma/\sigma\eq
\Sigma$. One can show that every di-analytic structure on $\Sigma$ gives a 
unique conformal structure on $\hat \Sigma$. The Teichm\"uller space 
$\calt(\Sigma)$ can be canonically identified with the fixed point set of 
$\calt(\hat \Sigma)$ under the involution $\sigma^*$:
  $$ \calt(\Sigma) \cong \calt(\hat \Sigma)_{\sigma^*} \, . $$
One can show that this space is real-analytically isomorphic to 
$\reals^{-3\chi}$ for Euler characteristic $\chi(\Sigma)\,{\leq}\,{-}1$, 
isomorphic to $\reals$ for Euler characteristic 0, and to a point for
$\chi(\Sigma)\eq1$. It can also be shown that Teichm\"uller spaces for
surfaces $\Sigma$ with the same Euler characteristic are real-analytically
isomorphic.

Finally, using the action of the mapping class group $\Omega(\Sigma)$
on $\calt(\hat\Sigma)$ one can canonically identify $\Omega(\Sigma)$ 
with the commutant $\Omega_{\sigma}(\hat\Sigma)$ of $\sigma^*$ in 
$\Omega(\hat\Sigma)$. This subgroup
is also called the {\em relative modular group\/}. We can therefore
identify the moduli space $\calm(\Sigma)$ with the quotient
  \be \calm(\Sigma) = \calt(\hat \Sigma)_{\sigma^*} /
  \Omega_{\sigma}(\hat\Sigma) \, . \ee 

\subsection{First remarks on strings}

We are now in a position to describe briefly the central goal of (perturbative)
string theory: Construct consistent quantum field theories by formulating
a perturbation expansion as a sum over \twodim\
manifolds, rather than use (as in Lagrangian quantum field theory)
Feynman diagrams, which are graphs, i.e.\ singular
one-dimensional manifolds. What is truly remarkable about string theory is
that this idea is applicable even for theories that include gravity
and do not possess a well-behaved Feynman perturbation expansion,
see e.g.\ \cite{POlc}.

In the same way we would not assign any physical reality to the lines of
a Feynman diagram, we should not directly assign a physical meaning to
the string world sheet. As a consequence, a basic theme of string theory
can be formulated as ``get rid of the world sheet''. This is done by
taking first the cohomology with respect to the (super-)Virasoro algebra,
then one eliminates the conformal structure by integrating over it, and 
finally, one sums over all genera, weighted by the string coupling constant
${\rm g}_{\rm str}$, to even get rid of topological aspects of the world sheet. 
Yet, after all these manipulations the string theory one obtains still 
strongly depends on the \cft\ model one started with.

If string theory indeed provides a perturbative quantization of certain 
classical field theories including gravitation, then the following question 
arises. These field theories possess non-trivial classical solutions 
\cite{Dukl}. Among these solutions there are configurations that generalize 
the Schwarzschild solution of Einstein gravity in the sense that they not 
just describe a black hole, i.e.\ a singularity of signature $1\,{+}\,0$,
but also higher-dimensional singularities of signature $1\,{+}\,p$, 
so-called black $p$-branes.

In certain cases  it is known how to set up ordinary field theoretic 
perturbation theory around such solitonic solutions, e.g.\ around instanton 
solutions of four-dimensional Yang\hy Mills theory. It was a major 
breakthrough for string theory when Polchinski discovered a prescription for
{\em string\/} perturbation theory in the background of $p$-branes.
His idea was to use open strings that are constrained to end on the
black brane. This idea triggered enormous progress in
the understanding of (some) non-perturbative sectors of string theory.
For us, this constitutes a strong motivation to study the corresponding 
world sheet theories, that is, \cft\ on surfaces with boundaries.  
String theory also has other classical solutions than D-branes, like
the Neveu\hy Schwarz five-brane or the fundamental string solution. 
For the five-brane a string perturbation theory has been proposed
in terms of a different chiral CFT than the one of
the flat background. This proposal is rather different in structure from the
one for string perturbation theory in the background of D-branes, where
the same chiral CFT as for the configuration without branes is used.  

\section{Algebraic aspects of chiral CFT: Vertex operator algebras}

This section is devoted to the study of algebraic and representation
theoretic aspects of chiral CFT. The main notion we will introduce
is the one of a vertex operator algebra (VOA). It formalizes the
notion of a chiral symmetry algebra.\,%
 \footnote{~Unfortunately, the term chiral algebra has been used in the
 mathematics literature for different, though related, objects, see 
 e.g.\ \cite{gaiT}.}

\subsection{The Virasoro algebra and its super-extensions}

First, however, we need a few facts about infinitesimal conformal
symmetries in two dimensions and its super-extensions. 
Conformal symmetry is encoded in the {\em Virasoro algebra\/}. This is 
the central extension of the Lie algebra of vector fields on a circle 
$S^1$; thus it is the \infdim\ \lie\ that is spanned by generators $L_n$ 
with $n\in\zet$ and a central element $C$, subject to the relations
  \be\begin{array}{lll}
  [L_n,L_m] &=& (n\,{-}\,m)\, L_{n+m} + \frac1{12}\,(n^3\,{-}\,n)\,
  \delta_{n+m,0}\, C \,,  \\[.43em]  {}[L_n,C] &=& 0 \,.  
  \end{array}\labl{vira} 
If $v$ is an eigenvector of $L_0$ in a representation of the Virasoro algebra, 
then its eigenvalue is called
the {\em conformal weight\/} of $v$ and denoted by $\Delta_v$.
Using a formal variable $z$, we can combine the generators into a `field'
  $$ T(z) = \sum_{n\in\zet} L_n\, z^{-n-2} \,,  $$
called the chiral stress-energy tensor. 
The formal variable $z$ allows us to characterize the Virasoro algebra
also as a central extension of the Lie algebra of derivations of the
field of rational functions in $z$.

There are various super-extensions of this construction. The simplest one,
the $N\eq1$ superconformal algebra (or $N\eq1$ Virasoro algebra),
is described by the superfield
  $$ {\cal T}(z,\vartheta) = \Frac12\, G(z) + \vartheta\, T(z) \,,  $$
whose first component $G(z)$ has the expansion
  $$ G(z) = \sum_{r\in\zet+\eps} G_r\, z^{-r-3/2} \,.  $$
The parameter $\eps\iN\{0,\frac12\}$ depends on the `sector': 
$\eps\eq0$ in the {\em Ramond\/} sector, while
$\eps\eq1/2$ in the {\em Neveu\hy Schwarz\/} sector.
For later purposes one should keep in mind that Ramond and Neveu\hy Schwarz
sector are distinguished by the monodromies of a specific field. 
The commutation relations among $G$ and $T$ read
  $$ \begin{array}{lll}
  [L_n, G_r] &=& (\frac12 n \,{-}\, r)\, G_{n+r} \,,  \\[.63em]
  \{G_r,G_s\} &=& 2L_{r+s}+\frac13(r^2{-}\,\frac14)\,\delta_{r+s,0}\, C \,. 
  \end{array}$$
The $N\eq1$ Virasoro algebra also possesses a geometric interpretation. Consider
the punctured superdisc $(\complex^{\times\!})^{1|1}$. This space is characterized
by the space of functions on it, which is $\complex[z,z^{-1},\vartheta]$, i.e.\
consists of Laurent polynomials in a usual bosonic variable $z$ and a Grassmann
variable $\vartheta$. It possesses a {\em contact structure}, given by
$\rmd z\,{-}\,\vartheta\,\rmd\vartheta$, and the Lie algebra of derivations 
of $\complex[z,z^{-1},\vartheta]$ that preserve this contact structure is 
precisely the $N\eq1$ Virasoro algebra.

For the applications we have in mind we need a non-minimal supersymmetric 
extension of the Virasoro algebra -- the $N\eq2$ {\em superconformal 
algebra\/} (or $N\eq2$ Virasoro algebra), which apart from $T(z)$ 
contains two supercurrents $G^\pm(z)$ as well as an abelian current $J(z)$.
Their non-vanishing $\mbox{(anti-)}$\,commutation relations read
  \be \begin{array}{lll}
  \{G^+_r,G^-_s\} &=& 2L_{r+s} + (r\,{-}\,s)\, J_{r+s} + \frac13(r^2{-}\,
  \frac14)\, \delta_{r+s,0}\, C \,,  \\[.63em]
  {}[J_n,J_m] &=& \frac13\, \delta_{n+m,0}\, C \,,  \\[.63em]
  {}[J_n, G^\pm_r] &=& \pm\, G^\pm_{n+r} \,.
  \end{array}\ee
It will be important that this algebra admits a family of automorphisms, 
indexed by a parameter $\theta\iN\frac12\zet$ and known as 
{\em spectral flow\/}. These automorphisms act as
  $$\begin{array}{lll}
  \omega_\theta(L_n) &=& L_n-\theta J_n+\frac C6\,\theta^2\,\delta_{n,0}
  \,, \\[.53em]
  \omega_\theta(J_n) &=& J_n - \frac C3\, \theta\, \delta_{n,0} \,, \\[.53em]
  \omega_\theta(G^\pm_r) &=& G^\pm_{r\mp\theta} \,.
  \end{array}$$

The $N\,{=}\,2$ super Virasoro algebra also admits another family of
automorphisms, called {\em odd spectral flow\/}. It is again indexed
by a parameter $\theta\iN\frac12\zet$:
  $$\begin{array}{lll}
  \alpha_\theta(L_n) &=& L_n+\theta J_n+\frac C6\,\theta^2\,\delta_{n,0}
  \,, \\[.53em]
  \alpha_\theta(J_n) &=& - J_n - \frac C3\,\theta\, \delta_{n,0} \,, \\[.53em]
  \alpha_\theta(G^\pm_r) &=& G^\mp_{r\mp\theta} \,.
  \end{array}$$
While the ordinary spectral flow provides a group of automorphisms,
satisfying $\omega_{\theta_1}{\circ}\,\omega_{\theta_2}\eq
\omega_{\theta_1+\theta_2}$, the odd spectral flow obeys the composition law
$\alpha_{\theta_1}{\circ}\,\alpha_{\theta_2}\eq\omega_{\theta_2-\theta_1}$; 
in particular all $\alpha_\theta$ are involutions.
The two types of flows are related by 
  $$  \alpha_{\theta_2}^{} {\circ}\, \omega_{\theta_1}^{}
  = \alpha_{\theta_2-\theta_1}^{} \qquad{\rm and}\qquad
  \omega_{\theta_2}^{} {\circ}\, \alpha_{\theta_1}^{} = \alpha_{\theta_1+\theta_2}^{}
  \,. $$

\subsection{Chiral algebras}

We are now ready to introduce the notation of a chiral algebra. The following
structure is expected to be present in every chiral CFT: First,
a {\em vacuum vector\/} $v_\Omega$ and its `descendants', which span a vector 
space $\calh_\Omega$; they should somehow encode the chiral symmetries of the 
theory. And second,  a {\em field-state correspondence\/} $Y$ which
associates fields to states in a such way that applying the field to the
vacuum $v_\Omega$ gives back the state -- a structure familiar from general
quantum field theory.

All these ideas are formalized in the notion of a {\em vertex operator 
algebra\/} ({\em VOA\/}). A VOA consists of the following data:
\nxt a space of states -- a graded vector space 
  $$ \calh_\Omega = \bigoplus_{n=0}^\infty \calh_{(n)} $$
     {}$\;\ $~whose homogeneous subspaces $\calh_{(n)}$ are \findim,
     $\dim\calh_{(n)}{<}\,\infty$;
\nxt a vacuum vector $v_\Omega\iN\calh_\Omega$;
\nxt a shift operator 
  $$T: \quad \calh_\Omega \to\calh_\Omega \,; $$
\nxt and a field-state correspondence $Y$ involving a formal variable $z$:
  \be  Y:\quad \calh_\Omega \to {\rm End}(\calh_\Omega)[[z,z^{-1}]] \,;  \ee
$Y(v,\cdot)$ is also called the {\em vertex operator\/}
for the vector $v\iN\calh_\Omega$.
\\[.25em]
\mbox{}$\;\ $~These data are subject to the conditions
\nxt that the field for the vacuum $v_\Omega$ is the identity,\,
     $Y(v_\Omega,z)\eq\id_{\calh_\Omega}$; 
\nxt that the field-state correspondence respects the grading, i.e.\ if
     $v\iN\calh_{(n)}$, then \\ \mbox{}$\;\ $~all endomorphisms $v_m$ appearing in 
     $Y(v,z)\eq\sum_{m\in\zet} v_m z^{-n-m}$ have grade \\ \mbox{}$\;\ $~$m$:
     $v_m (\calh_{(p)}) \,{\subseteq}\, \calh_{(p+n)}$; 
\nxt that one recovers states by acting with the corresponding fields on the
     va-\\ \mbox{}$\;\ $~cuum and `sending $z$ to zero', or more precisely,
  $$ Y(v,z) v_\Omega\, \in\, v + z\, \calh_\Omega[[z]]  $$
     {}$\;\ $~for every $v\iN\calh_\Omega$;
\nxt that $T$ implements infinitesimal translations,
  \be  [T,Y(v,z)] = \partial_z Y (v,z) \,;  \ee
     {}$\;\ $~and that the vacuum is translation invariant, $\,Tv_\Omega\eq0$.
\nxt Finally, the most non-trivial constraint -- called {\em locality\/} --
     is that commuta-\\ \mbox{}$\;\ $~tors of fields have poles of at most
     finite order. More precisely, for any
     two\\ \mbox{}$\;\ $~$v_1,v_2\iN\calh_\Omega$ there must exist a number
     $N\eq N(v_1,v_2)$ such that
  \be  (z_1\,{-}\,z_2)^N\, [Y(v_1,z_1), Y(v_2,z_2) ] = 0 \,. \labl{2.2}
This is a kind of weak commutativity. Note that this constraint only makes 
sense because we consider formal series in the $z_i$, which can extend
to both arbitrarily large positive and negative powers. Had we restricted
ourselves to ordinary Laurent series, i.e.\ series without arbitrarily
large negative powers, \erf{2.2} would already imply that the commutator
vanishes.  

We have motivated the structure of a VOA by physical requirements on a
chiral symmetry algebra. There have also been attempts to understand this 
structure from a purely mathematical point of view, by regarding VOAs as 
(singular) rings in suitable categories. The aim of these constructions has
been to better understand quantum affine \lie s \cite{soib4} and to obtain a
generalization of VOAs for conformal field theories in higher dimensions
\cite{borc11}; these issues are beyond the scope of the present lectures.

For the application to CFT one needs a special class of VOAs, namely
{\em conformal\/} VOAs. A VOA is called conformal,
of central charge $c$, if there exists a complex number $c$ and
a vector $v_{\Vir}\iN\calh_{(2)}$ such that operators appearing in the
mode expansion of
  $$ T(z) := Y(v_{\Vir},z) = \sum_{n\in\zet} L_n\, z^{-n-2} $$
possess the following properties. $ L_{-1}\eq T$ gives the translations;
$L_0$ is semisimple and reproduces the grading of $\calh_\Omega$;\,%
 \footnote{~This axiom excludes so-called logarithmic CFTs
 from our considerations.}
$L_0$ acts as $n\,\id$ on $\calh_{(n)}$; and finally,
  $$ L_2 v_{\Vir} = \Frac12\, c\, v_\Omega \,. $$
These axioms imply that the modes $L_n$ span a Virasoro algebra of central 
charge $c$.

Let us also mention that the choice of $v_{\Vir}$ for a given VOA and a 
given shift operator $T$ is not necessarily unique; it is sometimes also 
called the choice of a `conformal structure'. This should not be confused 
with the conformal structure of the world sheet; it is, rather, related to 
(chiral) properties of the target space.

For string theory, we will need VOAs with even more structure. We start
with the notion of a {\em graded VOA\/}, which is a VOA with an additional
grading over $\zet$:
  $$ \calh_\Omega = \bigoplus_{n=0}^\infty \mbox{\Large(} \bigoplus_{g\in\zet}
  \calh_{(n)}^{(g)} \mbox{\Large)}  \,.  $$
We call $g$ the {\em ghost number\/} and require that the field-state 
correspondence $Y$ has ghost degree 0. We would also like to have an 
operator on $\calh_\Omega$ whose eigenvalue gives the ghost number. Rather 
than to introduce  that operator directly, we will adhere to the following
principle: Avoid introducing operators on $\calh_\Omega$ by hand, rather
obtain them via the field-state correspondence from states in $\calh_\Omega$.
Thus for a graded VOA we require in addition that there is a state
$v_F\iN\calh_{(1)}^{(0)}$ such that the eigenvalues of
  $$ F_0 := {\cal R}es_{z=0} Y(v_F,z) $$
gives the ghost number. The field $Y(v_F,z) $ is also called ghost number
current.

A particularly important case of graded VOAs are topological VOAs, 
TVOAs for short. Again we introduce more structure by requiring the
existence of vectors in $\calh_\Omega$: We demand that there are
$v_B\iN\calh_{(2)}^{(-1)}$ and $v_G\iN\calh_{(1)}^{(1)}$ such that
\nxt $Q\,{:=}\, {\cal R}es_{z=0} Y(v_G,z)$ is nilpotent, $Q^2\eq0$. \\
     $Q$ is called BRST charge, and the field $Y(v_G,z)$ is called BRST
     current. 
\nxt The stress-energy tensor is cohomologically trivial,
  \be \{ Q, Y(v_B,z)\} = Y(v_{\Vir},z) \equiv T(z) \,.  \labl{tvir}

A few statements about TVOAs are immediate consequences of these definitions:
\nxt It follows from \erf{tvir} that the Virasoro algebra is $Q$-exact, 
     $[Q,L_n]\eq0$. Together with the Jacobi identity and 
     the nilpotency of $Q$ this implies that
  $$ [L_n,L_m] = [L_n,\{Q,b_m\}] = \{Q,[L_n,b_m]\} = (n-m) \{Q,L_{n+m}\}
  = (n-m) L_{n+m} , $$
     i.e.\ the Virasoro algebra has vanishing central charge. 
\nxt The nilpotency of $Q$ allows us to define its cohomology $H_Q$.
     Translations in $z$ are represented by $L_{-1}$, which is cohomologically
     trivial; as a consequence, correlators of closed states do not depend
     on $z$. Thus the cohomology $H_Q$ carries the structure of a \twodim\
     topological field theory.

It is easy to give examples of TVOAs. Every $N\eq2$ VOA gives
in fact rise to two different TVOAs. Indeed,
the vectors $(L_{-2} \,{\pm}\,\frac12 J_{-2})v_\Omega$ provide two different
conformal structures on $\calh_\Omega$ with vanishing central charge.
The corresponding stress-energy tensors read
  $$ T^\pm(z) = T(z) \pm \Frac 12\, \partial J(z) \,; $$
this `modification' of $T(z)$ is also known as a {\em topological twisting\/}.
With respect to $T^+$, the supercurrent $G^+$ has conformal weight 1
and provides the BRST current, while $G^-$ has conformal weight $2$
and gives the antighost current; with respect to $T^-$, the roles of
$G^+$ and $G^-$ are interchanged. The ghost number grading is provided by the 
U(1) current $\pm J(z)$.

We conclude this section with a word of warning. The notion of a VOA
is a mathematical formalization of a physical concept, the algebra of
chiral symmetries. It is, however, {\em not\/} the only available formalization
of that physical concept. For instance, there is another formalization
which treats $z$ as a complex number rather than as a formal variable
\cite{gago}; in this framework one attempts to reconstruct 
the chiral algebra from the values of the $n$-point blocks on a 
\findim\ subspace of $\calh_\Omega$. A rather different approach, based on 
the notion of local observables \cite{HAag}, describes the chiral symmetries
in terms of nets of von Neumann algebras over a circle $S^1$ (for 
details see e.g.\ \cite{EVka}). The relation between these different
incarnations of chiral symmetries remains to be clarified.

\subsection{Examples}

It is time for presenting examples of VOAs. We start with the chiral CFT
for a single free boson; it is based on the Heisenberg algebra, which has
generators $b_n$ with $n\iN\zet$ and relations
  $$[b_n,b_m] = n\, \delta_{n+m,0} \, . $$
In this case $\calh_\Omega$ is nothing but a Fock space, and $v_\Omega$ is 
the ground state in this Fock space. To define the field-state correspondence
one introduces abelian currents
  $$ J(z) = \sum_{n\in\zet} b_n\, z^{-n-1} $$
and identifies $Y(b_{-1} v_\Omega,z)\eq J(z)$. More generally, one sets
  $$ Y(b_{n_1}{\cdots}\, b_{n_k}v_\Omega,z) = \frac1{(n_1{-}1)!\cdots(n_k{-}1)!}
  \,\mbox{\bf:}\, \partial_z^{n_1-1}J(z) \cdots\, \partial_z^{n_k-1}J(z)\,
  \mbox{\bf:} \,, $$
where the colons indicate a normal ordering. 
This prescription indeed yields the structure of a VOA.
(It is not a trivial exercise, though, to check that this works out.)

This example has an important generalization. Let $L$ be a lattice, and
$V\eq L\,{\otimes}_{\zet}^{}\reals$ be the associated real vector space with 
basis $\{b^{(i)}\}$. Suppose that $V$ has a non-degenerate bilinear form 
$\kappa$. To the \infdim\ \lie\ with basis $b^{(i)}_n$, $n\iN\zet$, and 
relations
  $$ [b^{(i)}_n, b^{(j)}_m] = n\, \kappa(b^{(i)},b^{(j)})\, \delta_{n+m,0} $$
one associates a Fock space $\calh_\Omega$. One checks that it carries again
the structure of a VOA. This structure can be further generalized: Suppose
that the bilinear form is such that $L$ is an even lattice, i.e.\
$\kappa(v,w)\iN\zet$ and $\kappa(v,v)\iN2\zet$ for all $v,w\iN L$. Then
the space
  $$ \calh_\Omega \,{\otimes}\, \complex[L] \,, $$
where $\complex[L]$ is the group algebra of $L$, has the structure of a VOA, 
too. It is called the lattice VOA for $L$ and describes $\dim V\eq{\rm rank}\,L$
many compactified chiral free bosons. 

Similar constructions are possible when choosing other \infdim\ \lie s in 
place of the Heisenberg algebra. One can e.g.\ take the Virasoro algebra 
itself; then $T(z)$ roughly plays the role of the abelian current $J(z)$. 
When the chiral algebra is generated solely from the Virasoro algebra,
then the model is called a Virasoro {\em minimal model\/}.\,%
 \footnote{~The term `minimal model' is sometimes also used for rational
 chiral CFTs whose modules are finitely reducible in terms of modules over the
 Virasoro algebra.}
Another important class of examples is furnished by untwisted affine \lie s, 
where non-abelian currents $J^a(z)$ ($a$ an adjoint label of the underlying 
\findim\ simple \lie) take over the role of $J(z)$; the models obtained this 
way are known as Wess\hy Zu\-mi\-no\hy Wit\-ten ({\em WZW\/}) models. 
The Heisenberg, Virasoro and affine \lie s belong to the so-called Lie algebras 
of formal distributions; these are Lie algebras $\g$ that are spanned over
$\complex$ by the coefficients of a collection of $\g$-valued mutually
local formal distributions $\{a^\alpha(z)\}$. 

The previous construction of a VOA for the free boson can be generalized
to arbitrary such Lie algebras $\g$. Namely, to $\g$ together 
with a vector space endomorphism $T$ of $\g$ such that 
$Ta^\alpha(z)\eq\partial a^\alpha(z)$ and a weight on the positive subalgebra
of $\g$, one can construct in a canonical way a VOA.\,%
 \footnote{~In the context of these VOAs, the term `chiral algebra' is
 sometimes also used to refer to the Lie algebra of formal distributions, 
 rather than to the VOA itself.}
For more information, we refer to sections 2.7 and 4.7 of \cite{KAc4}.

Our last example are the so-called {\em first order systems\/}. This
is a family of VOAs, labelled by two parameters $\lambda\iN
\zet/2$ and $\eta\eq{\pm}1$. One starts with a Lie algebra generated by
two formal distributions
  $$  b(z) = \sum_{n\in\epsilon+\zet} b_n z^{-n-\lambda} \,, \qquad
  c(z) = \sum_{n\in\epsilon+\zet} c_n z^{-n-(1-\lambda)} $$
of conformal weight $\lambda$ and $1\,{-}\,\lambda$, respectively. These fields
are bosonic for $\eta\eq{-}1$ and fermionic for $\eta\eq1$. In the former 
case, $\epsilon$ is zero, while in the latter it
takes the values $0$ for the Ramond sector and $1/2$ for the
Neveu\hy Schwarz sector. The modes of $b$ and $c$ obey the (anti-)commutation
relations $\{c_m,b_n\}_\eta^{}\eq\delta_{n+m,0}$. The VOA is defined on
a Fock space, built on a highest weight vector $v_\Omega$ with relations
  $$ b_n v_\Omega = 0 \ \mbox{ for } n\,{\ge}\,1{-}\lambda \,, \qquad
  b_n v_\Omega = 0 \ \mbox{ for } n\,{>}\, \lambda \, . $$ 
As the stress-energy tensor one takes
  $$T(z) = -\lambda\, \mbox{\bf:}b\,\partial c\mbox{\bf:} - (1{-}\lambda)\,
  \mbox{\bf:}\partial b\, c\mbox{\bf:}\,; $$
the Virasoro central charge is $c\eq1{-}3Q^2$ with
$Q\,{:=}\,\epsilon (1{-}2\lambda)$. There is a U(1) current with modes
$j_n\eq\sum_{m\in\zet+\epsilon} \mbox{\bf:}c_{n-m} b_m\mbox{\bf:}$ 
which is, however, anomalous:
  $$ [L_n,j_m] = \Frac12\, Q\, n(n{+}1)\, \delta_{n+m,0} - m\, j_{n+m} \, . $$
First order systems are of particular interest for the following values:
$\lambda\eq2,\,\eta\eq1$ yields the ghosts for bosonic reparametrizations of 
the string world sheet, $\lambda\eq3/2,\,\eta\eq{-}1$ gives the ones for 
gauging the fermionic operators of an $N\eq1$ superconformal symmetry on the 
world sheet, and $\lambda\eq1/2$, $\eta\eq1$ is just a complex free fermion.

We finally mention a recent
development in mathematics, the construction of the {\em chiral de Rham
complex\/} \cite{masv}. This allows to associate a vertex operator algebra
to every complex variety. For many more details about VOAs, see e.g.\ the 
recent review \cite{freN5}, which we have followed in this section.  

\subsection{Representations}

Recall that the VOA is formalizing aspects of a chiral symmetry
algebra, and symmetries in a quantum theory should be represented on 
the space of states. We are thus lead to study the representation theory
of VOAs. 

A {\em \rep\/} of a VOA is a graded vector space 
  $$ M = \bigoplus_{n\in\zet_{\geq0}} M_{(n)} $$
{}$\;\ $~with 
\nxt a translation operator $T_M{:}\; M\,{\to}\, M$ of degree 1 and
\nxt a representation map
  $$ Y_M:\quad \calh_\Omega \to {\rm End}(M)[[z,z^{-1}]] \,, $$
{}$\;\ $~such that
\nxt for $v\iN\calh_{(n)}$ all components of $Y_M(v,z)$ are endomorphisms
     of $\calh_\Omega$ of degree \\\mbox{}$\;\ $~$-n$;
\nxt $Y_M(v_\Omega)= \id_M\,;$
\nxt $[T_M,Y_M(v,z)] = \partial_z Y_M(v,z)\,; $
\nxt and 
  \be Y_M(v_1,z_1)\, Y_M(v_2,z_2) = Y_M( Y(v_1,z_1{-}z_2)v_2, z_2) \,. \labl{2.4}

One can show that the VOA furnishes a \rep\ of itself; this is called
the {\em vacuum representation\/}. This implies that the identity \erf{2.4}
is in particular valid for $\calh_\Omega$, i.e.\ \erf{2.4} remains true when
$Y_M$ is replaced by $Y$. This expresses a kind of associativity of the VOA.
Thus for VOAs `associativity' in the sense of \erf{2.4} is a consequence of
`commutativity' in the sense of \erf{2.2}. 

Once we know what a representation is, we can consider equivalence classes
of isomorphic \irrep s. They form a set $I$. We will call the elements
of $I$ {\em labels} or also, by an abuse of terminology, {\em primary
fields\/} or {\em sectors\/}. We use lower case greek letters for 
elements in $I$, and denote by $\calh_\mu$ the underlying vector space for 
a representation isomorphic to $\mu\iN I$.

When the VOA is conformal, every module $M$ is in particular, by restriction,
a module over the Virasoro algebra \erf{vira}. It follows directly from the 
definition of a conformal VOA that in each CFT model the central element $C$ 
of the Virasoro algebra acts as $C\eq c\,\id$ with one 
and the same value of the number $c$ in every \irrep\ that occurs in the model. 
Also recall that when $v$ is an eigenstate of the Virasoro zero mode
$L_0$, its eigenvalue $\Delta_v$ is called the {\em conformal weight\/} of $v$. 
The conformal weights of different vectors in the same irreducible module 
differ by integers, or in other words, $\eE^{2\pi\ii L_0}$ acts as a multiple 
of the identity on every irreducible module. We will therefore refer, somewhat
abusing terminology, to the {\em fractional part\/}
of the conformal weight of any eigenvector of $L_0$ in an irreducible 
module $M$ as the conformal weight $\Delta_M$ of the module $M$.

A vertex operator algebra is called {\em rational\/} if every module
is fully reducible. Rational VOAs possess two surprising properties 
\cite{dolm3}: The homogeneous subspaces $M_{(n)}$ of the \rep\ spaces $M$ are 
\findim, and there are only finitely many inequivalent \irrep s.
Notice, however, that not every VOA that has only finitely many inequivalent
irreducible representations is rational in this strong sense. For instance,
logarithmic CFTs have reducible, but not fully reducible modules, although
there may be only finitely many inequivalent irreducible modules. An example 
of a non-rational VOA is provided by the one
for a single free boson that has been discussed in Section 3.3. It has 
infinitely many inequivalent \irrep s, one for each real number $q$.
It also has a reducible but not fully reducible modules. Namely, start with a
\findim\ vector space $V$ on which $b_0$ has non-trivial Jordan decomposition
and cannot be diagonalized. It gives rise to a module over the Heisenberg 
algebra in the same way as the ordinary Fock space is generated from a \onedim\
space. This module is also a module of the VOA, and it is reducible, but
not fully reducible.

The above results on rational VOAs imply that for rational VOAs we can define 
the character of a representation as a formal function in $\tau$:
  \be  \chii_M(\tau) := {\rm tr}_M^{} \exp(2\pi\ii\tau(L_0\,{-}\,c/24)) \, . \ee
Another remarkable result \cite{zhu3} is that under a certain
(technical) finiteness condition
the characters are holomorphic functions in $\tau$ on the complex
upper half-plane $H$. Moreover, the finite set of characters of all
inequivalent irreducible modules carries an action of ${\rm SL}(2,\zet)$.

Quite generally, it is expected that the set of representations of a
rational VOA carries even more structure, namely the one of 
a {\em modular tensor category\/}. This formalizes the tensor product
of modules as well as the so-called Moore\hy Seiberg data like braiding,
fusing and the modular transformations of the characters. For more details
about modular tensor categories, we refer to \cite{BAki}.

\subsection{Remarks on models with extended supersymmetry}

We finally discuss some specific aspects of models with extended ($N\eq2$)
superconformal symmetry. (For a review, see also \cite{warn2}.) Unitary 
theories of this type exist for Virasoro central charges $c\,{\ge}\,3$ and 
for $c\eq 3k/(k+2)$ with $k$ a positive integer. The simplest realization 
with central
charge $c\eq3$ is constructed from two real chiral bosons $X^1(z),\,X^2(z)$ 
and two real chiral fermions $\psi^1(z),\,\psi^2(z)$. We may think of 
these fields as (su\-per-)coordinates of a target space. We endow this target 
space with a complex structure and consider complex fields
  $$ X^{\pm} = \Frac1{\sqrt2}\,(X^1{\pm}\,\ii X^2) \;\ \mbox{ and } \;\
  \psi^{\pm} = \Frac1{\sqrt2}\,(\psi^1{\pm}\,\ii \psi^2) \,. $$
The generators of the $N\eq2$ algebra can be expressed through these fields as
  $$ \begin{array}{l}
  T(z) = - \,\mbox{\bf:} \partial_z X^+\partial_z X^-(z) \mbox{\bf:}
  - \frac12 \left(  \mbox{\bf:} \partial\psi^+\psi^-(z) \mbox{\bf:}
  + \mbox{\bf:}  \partial\psi^-\psi^+(z) \mbox{\bf:}  \right) \,, \\[.52em]
  G^+(z) = -\frac12\, \psi^+ \partial_z X^-(z) \,,  \qquad\quad
  G^-(z) = -\frac12\, \psi^- \partial_z X^+(z) \\[.52em]
  J(z) = \mbox{\bf:} \psi^-\psi^+(z) \mbox{\bf:}  \,.  \end{array} $$
This example is easily generalized to $2k$ bosons and fermions. Any
complex structure on a real $2k$-dimensional vector space then gives
a realization of the $N\eq2$ algebra with central charge $c\eq3k$.

Our next example are $N\eq2$ {\em minimal models\/}.
They are similar to the Virasoro minimal models, whose
chiral algebra is generated by the Virasoro algebra
alone. Namely, the chiral algebra of an $N\eq2$ minimal model is generated 
by the bosonic subalgebra of the $N\eq2$ superconformal algebra. 
(Here it is essential that we regard the chiral algebra as a VOA, and not
as a so-called super-VOA.)

The $N\eq2$ minimal models come in a series, and theories are parameterized
by a non-negative integer $k$, the level. Their Virasoro central charge
is $c\eq 3k/(k{+}2)$. They can be realized by a coset construction,
  $$ {\mathfrak su}(2)_k \oplus {\mathfrak u}(1)_4 \,/\,
  {\mathfrak u}(1)_{2k+4} \,,  $$
and accordingly the \irrep s can conveniently be labelled as 
$\Phi^{l,s}_m$, with $l\iN\{0,1,...\,,k\}$, $s\iN\zet/4\zet$ and
$m\iN\zet/(2k{+}4)\zet$. (Only triples with $l\,{+}\,s\,{+}\,m\iN2\zet$ 
are allowed, and $\Phi^{l,s}_m$ and $\Phi^{k-l,s+2}_{m+(k+2)}$ refer to
isomorphic \irrep s.)

We call a state $v$ in the NS sector $N\eq2$\,-primary if there are numbers
$\Delta$ and $q$ such that
  $$ \begin{array}l
  G_r^\pm v = 0  \quad \mbox{ for }\; r\,{\ge}\,1/2 \,, \\[.4em]
  L_n v = J_n v = 0 \quad \mbox{ for }\; n\,{\ge}\,0 \,, \\[.4em]
  L_0 v = \Delta v  \quad \mbox{ and } \quad J_0 v = q v \,.
  \end{array} $$
{\em Chiral primaries} are $N\eq2$\,-primary states which in addition obey
$G_{-1/2}^+ v\eq0$, while anti-chiral primaries obey $G_{-1/2}^- v\eq0$.
The reader should be warned that the qualification ``chiral'' is meant here 
in a different sense than in the term chiral CFT; it refers to the fact that
a state is annihilated by half of the supersymmetry charges.

In unitary $N\eq2$ theories the identity
  $$ 0 \leq \langle v,\{G_{-1/2}^\pm,G^\mp_{1/2}\} v\rangle =
  (2\Delta_v{\mp}q_v)\, \langle v,v\rangle $$
holds; this gives immediately the unitarity bound
  $$ \Delta_v \ge \Frac12\, |q_v| \,. $$
States with positive charge $q_v$ saturate this bound if and only if
they are chiral primaries; similarly, anti-chiral primaries are precisely
the states of negative charge saturating the bound. Analogous arguments
show that in the Ramond sector the conformal weight is bounded from below by 
$c/24$; states satisfying this bound are called {\em Ramond ground states\/}. 

It turns out that the space of all (anti-)chiral primary states is \findim;
each of these two spaces can be endowed with a product. by the prescription
  \be v_1 \star v_2 := {\calr}es_{z=0} ( \Frac1z\, Y(v_1,z) v_2) \,. 
  \labl{cprod}
Using standard properties of VOAs, one can show that this product
is associative and commutative. This way one obtains the so-called chiral 
and the anti-chiral ring; actually these are not rings, but rather algebras 
over the complex numbers. This algebra is not semi-simple; on the contrary,
owing to charge conservation in the product \erf{cprod}, all generators except 
for the identity are nilpotent. In the case of an $N\eq2$ minimal model at 
level $k$, the chiral ring is isomorphic to $\complex[x]/(x^{k+1})$, with
$x^l\,{\cong}\,\Phi^{l,0}_l$. One can show that spectral flow maps anti-chiral 
primaries to Ramond ground states, and the latter to chiral primaries. Finally 
we mention that there is a Hodge-type decomposition theorem: Every vector 
$v$ in an $N\eq2$ superconformal theory can be written in the form
  $$ v= v_0 + G_{-1/2}^+ v' + G_{1/2}^- v'' $$
with $v_0$ a chiral primary. Chiral primaries can therefore be thought of
as analogues of (one half of) harmonic forms.  

\section{Geometric aspects of chiral CFT: Conformal blocks}

We now to combine the subjects of the two previous sections,
VOAs and complex curves. This will lead us to the central notion
of a conformal block.

\subsection{Conformal blocks}

Conformal blocks formalize the following physical idea. To any vector
$v\iN\calh_\mu$, we would like to associate a ``field'' $\Phi(v,\cdot)$
and, given a complex curve $\hat X$ and $m$ non-coinciding points $p_i$
on $\hat X$, we would like to associate a ``correlator'' 
$\langle \Phi(v_1,p_1) \cdots\, \Phi(v_m,p_m)\rangle$ to products of such
fields. We have put the words ``field'' and ``correlator'' in quotation
marks, for, as we will see, the quantities we will obtain do not enjoy all
properties we usually expect for correlators in a local quantum field theory.

We want to characterize the ``correlators'' by {\em Ward identities\/}
which express the chiral symmetries globally on the curve $\hat X$. To write
down these identities in a compact form, it helps to adjust the notation
and to write the ``correlator'' as 
  $$ \langle \Phi(v_1,p_1) \cdots\, \Phi(v_m,p_m)\rangle
  = \beta_{\vec p,\hat X}(v_1\ot\cdots\ot v_m) \,,$$
i.e.\ interpret it as a linear functional
  $$ \beta_{\vec p,\hat X}\in {(\calh_{\lambda_1}{\otimes}\,{\cdots}\,
  {\otimes}\, \calh_{\lambda_m})}^* $$
that depends on the curve $\hat X$ and
on the positions $\vec p\eq(p_1,p_2,...\,,p_m)$ of the field insertions.

The Ward identities make use of
the {\em holomorphic\/} structure on $\hat X$. We restrict ourselves to
the case of \wzwts, where the VOA is generated by non-abelian currents
$J^a(z)$ that correspond to a \findim\ simple \lie\ $\bar\g$.
We first introduce the (associative, commutative) algebra $\calf$ of 
{\em holomorphic\/} functions on $\hat X{\setminus}\,\vec p$ which have at
most finite order poles at the points $p_i$. The corresponding algebra 
$\bar\g\,{\otimes}\,\calf$ of $\bar\g$-valued functions 
has an action on
  $$ \vec\calh_{\vec\lambda} := \calh_{\lambda_1}{\otimes}\, \cdots\,
  {\otimes}\, \calh_{\lambda_m} \, . $$
To describe this action, we choose local coordinates $\xi_i$ around the 
insertion points $p_i$. Given a homogeneous element 
$J^a\ot f\iN\bar\g{\otimes}\calf$, for each insertion point $p_i$ we 
consider the expansion of $f$ in a Laurent series in $\xi_i$,
  $$ f^{(i)}(\xi_i) = \sum_{n\gg -\infty} a_n^{(i)}\,\xi_i^n \,. $$
The coefficients $a_n^{(i)}\iN\complex$ yield an element 
  $$ \sum_{n\gg-\infty} a_n^{(i)}\, J^a_n $$
of the corresponding untwisted affine \lie\ $\g\eq\bar\g^{\sss(1)}$.
The $J^a_n$ are operators acting on the \irrep\ $\calh_{\lambda_k}$.
The action of $J^a\ot f$ on $\vec\calh_{\vec\lambda}$ is then
defined as
  $$ \sum_{i=1}^m \bfe\oT \cdots\oT
  \mbox{\Large(} \!\sum_{n\gg-\infty}\!\! a_n^{(i)}\, J^a_n \mbox{\Large)}
  \oT \cdots\oT \bfe \,, $$
where the non-trivial factor sits at the $i$th position.

The vector space of our interest consists of all vectors in 
$(\vec\calh_{\vec\lambda})^*_{}$ on which the dual action is trivial:
  \be  V_{\vec\lambda}(\hat X,\vec p) := {(\vec\calh_{\vec\lambda})}^*_0 \,. 
  \ee
This is called the space of {\em conformal blocks\/}.
Notice that the action of $\bar\g{\otimes}\calf$ depends on the choice of
local coordinates at the insertion points. Hence the conformal
blocks depend on that choice as well. For a conformal VOA, however, the
Virasoro algebra provides a natural action of the group of changes of local 
coordinates on the representation spaces. As a consequence, 
the conformal blocks transform covariantly under such choices. 

In all known rational chiral CFTs, in accordance with the Verlinde formula, 
the space $V_{\vec\lambda}$ is a \findim\ vector space; it is a subspace of
${(\vec\calh_{\vec\lambda})}^*$ that depends on the labels $\vec\lambda$, on
the curve $\hat X$, and the positions $\vec p$ of the insertion points. 
These vector spaces also turn out to be the spaces of physical states of 
certain three-dimensional topological field theories (TFTs), the 
Chern\hy Si\-mons theories. It must be emphasized, though, that this
relation between TFT and chiral CFT does by no means imply that the two 
theories are equivalent. Indeed, already the respective spaces of physical 
states are rather different. For chiral CFT, the state space is provided by 
the \infdim\ graded vector spaces $\calh_\lambda$ that underly the VOA-modules.
The spaces of conformal blocks are of independent interest in mathematics;
they provide nonabelian generalizations of theta functions, i.e.\ they
are naturally isomorphic to spaces of sections over moduli spaces 
of (stable equivalence classes of holomorphic) $G$-bundles over $\hat X$, 
where $G$ is the connected and simply connected complex Lie group whose Lie 
algebra is $\bar\g$. For more details, we refer to \cite{sorg5}.  

It is again time for an example. Take the sphere $\hat X\eq\complex{\dl P}^1$,
with the usual (quasi-)global coordinate $z$, and two insertions at $z_1\eq 0$ and
$z_2\eq\infty$. As local coordinates, we choose $\xi_1\eq z$ and $\xi_2\eq 1/z$.
The algebra $\calf$ is in this case the algebra of all polynomials 
in $z$ and $z^{-1}$, $\calf\eq\langle z^n,\, n{\in}\zet\rangle$. The element
$J^a\oT z^n$ acts via $J^a_n\ot\bfe\,{+}\,\bfe\ot J^a_{-n}$.
The two-point blocks are then functionals 
$\beta\iN{(\calh_\lambda{\otimes}\calh_\mu)}^*_{}$ with the property that
  \be \beta\circ(J^a_n\oT \bfe + \bfe \oT J^a_{-n} ) = 0 \labl{3.1}
for all $a\eq1,2,...\,,\dim\bar\g$ and all $n\iN\zet$. One can show that 
non-vanishing functionals $\beta$ obeying \erf{3.1} exist only if $\lambda$ 
and $\mu$ are conjugate $\bar\g$-weights, $\mu\eq\lambda^+_{}$. (An analogous 
notion of conjugate fields exists in every chiral CFT, see Subsection 4.3 below.)

Note that these linear functionals are in the {\em algebraic\/} dual of
$\calh_\lambda^{}{\otimes\,}\calh_{\lambda^+_{\phantom i}}$; the Hilbert space 
dual is too small to contain them. Still, one abuses bra-ket notation
and likes to write them as vectors of
$\calh_\lambda^{}\,{\otimes}\,\calh_{\lambda^+_{\phantom i}}$. In terms of 
these ``vectors'' $| B_\lambda\rangle$, formula \erf{3.1} is written as
  \be (J^a_n\oT \bfe + \bfe \oT J^a_{-n} )\, | B_\lambda\rangle 
  = 0 \,.  \labl{ishi}
The quantities $| B_\lambda\rangle$ show up in various circumstances. In 
the context of conformally invariant boundary conditions they are also 
known as {\em Ishibashi states\/}. The reader should keep in mind that these
are nothing but two-point blocks on the sphere. It is sometimes possible to 
write down the Ishibashi state $| B_\lambda\rangle$ explicitly; e.g.\ for 
theories based on a free boson, it can be written as a generalized coherent state,
  $$ | B_\lambda\rangle = \exp\mbox{\Large(}{-}\sum_{n=1}^\infty b_{-n}\oT b_{-n}
  \mbox{\Large)}\, v_\lambda \,, $$
where $v_\lambda$ is the highest weight state in the tensor product of
Fock spaces. Such a realization is helpful when one is interested in calculating
one-point functions on a disc explicitly. It is, however, {\em not\/}
necessary to know such an explicit realization if one wants to determine
the spectrum of boundary fields. In this case, it is sufficient to know
how $| B_\lambda\rangle$ behaves under factorization (see below).
The crucial information that allows to calculate concretely with boundary states
is the following identity that relates two-point blocks and characters:
  $$ \chii_\lambda(2\tau) = \langle B_\lambda |\, \eE^{2\pi\ii \tau( 
  L_0\ots\bfe + \bfe\ots L_0 - c/12)}\,| B_\lambda\rangle \, . $$

\medskip

We presented the definition of conformal blocks for specific
VOAs, the ones that underly \wzwts. A general approach has been developped 
in \cite{freN5}; here we just sketch the idea.
For \wzwts\ it was sufficient to take into account only the Ward
identities implied by the currents, i.e.\ by the Virasoro-primary fields of 
conformal weight $\Delta\eq 1$. In the general case we must implement
more information about the VOA. The idea is to define a bundle 
$\tilde\calh_\Omega$ over the curve $\hat X$ whose fibers are isomorphic to 
$\calh_\Omega$. Roughly speaking, one wants to have an appropriate copy of the 
VOA at each point of the curve. (The bundle $\tilde\calh_\Omega$ is actually 
twisted in such a way that covariance under changes 
of local coordinates is implemented; for simplicity, we suppress this 
point in our discussion.) Similarly one defines a bundle $\tilde\calh_\lambda$ 
over $\hat X$ for each module $\calh_\lambda$, again with the goal to have 
at one's disposal, as the fiber of $\tilde\calh_\lambda$ at $x\iN X$, a copy of 
the module over every point $\hat x$ of the curve $\hat X$.

The field-state correspondence $Y$ of the VOA then translates into the following
structure. Locally on a small disc around every point of the curve we have
a section $\caly$ in the dual bundle ${\tilde\calh_\Omega}^*$ with 
values in the endomorphisms of $\calh_\Omega$. This is the analogue to having
for each value of the formal coordinate $z$ a map that associates to
every vector $v\iN \calh_\Omega$ an endomorphism of $\calh_\Omega$.

The conformal blocks are defined as linear forms $\varphi$ on the tensor 
product of the vector spaces $\tilde\calh_{\lambda_i;p_i}$, i.e.\ the fibers of 
$\tilde\calh_{\lambda_i}$ over the insertion points $p_i$. Again we must select
the conformal blocks by an invariance property that uses global
{\em holomorphic\/} features of the curve $\hat X$. We require that 
for every $m$-tuple of vectors $v_i\iN\calh_{\lambda_i}$ the sections
  $$\varphi(v_1\ot\cdots\ot\,\caly(\cdot,\cdot)v_i\,\ot\cdots\ot v_m) $$ 
in the restriction of the bundle ${\tilde\calh_\Omega}^*$ over the local discs
around the points $p_i$ can be extended to a global holomorphic section
on the punctured curve $\hat X{\setminus}\{\vec p\}$.

One can apply the section $\varphi$ in ${\tilde\calh_\Omega}^*$ in particular 
to a Virasoro-primary field of conformal weight $\Delta\iN\zet_{\geq0}$. Currents 
are such fields with $\Delta\eq1$, and the resulting Ward identities are 
expressible in terms of functions, i.e.\ 
sections in a power $\calk^0\eq\calk^{1-\Delta}$ of the canonical line bundle 
of $\hat X$. One can show that for arbitrary $\Delta$ the general prescription
implies Ward identities that use sections in $\calk^{1-\Delta}$.

\medskip

We finally introduce the notion of a {\em tensor product} of two
chiral CFTs. This plays a crucial role in the construction of string
vacua. Its chiral algebra is a tensor product of two conformal VOAs
$(\calh_\Omega^{(i)}, v_\Omega^{(i)}, v_\Vir^{(i)}, T_i, Y_i)$.
Indeed, it is not hard to check that the data
  $$\begin{array}l
  \calh^{}_\Omega := \calh_\Omega^{(1)}\,{\otimes}\, \calh_\Omega^{(2)} \,,
  \qquad v_\Omega := v_\Omega^{(1)} \oT v_\Omega^{(2)} \,, \qquad
  T := T_1\oT\bfe + \bfe\oT T_2 \,,  \\[.66em]
  v_{\Vir}^{} := v^{(1)}_{\Vir}\oT v_\Omega^{(2)} +
  v_\Omega^{(1)}\oT v^{(2)}_{\Vir} \,, \qquad Y := Y_1\oT Y_2 
  \end{array} $$
define a new conformal VOA of central charge $c\eq c_1{+}\,c_2$.
Its \irrep s are tensor products of \irrep s,
$\calh_{\lambda_1}{\otimes}\,\calh_{\lambda_2}$, and the conformal blocks
are tensor products of vector spaces as well. 

We end with a word of warning. In the bosonic language we are using here, 
the tensor product of superconformal theories 
is {\em not\/} a superconformal theory any longer.
The reason for this is simple: The supercurrent $G$ is not a field in
the chiral algebra, since it does not have integral conformal weight.
Rather, it corresponds to a state $G_{-3/2} v_\Omega$ in some different
sector that we call $\calh_{\rm v}$. For the total supercurrent in a tensor
product of two superconformal theories, we would like to take
$G(z)\ot\bfe+\bfe\ot G(z)$. But this sum is not well-defined; indeed, the two 
terms of the sum are in two different superselection sectors $\calh_{\rm v}
{\otimes}\calh_\Omega$ and $\calh_\Omega{\otimes}\calh_{\rm v}$ of the 
tensor product.  We will see later how to deal correctly with this situation.

\subsection{Moduli spaces}

So far, we have kept the curve $\hat X$ and the insertion
points $\vec p$ fixed. Now we investigate what happens when they are varied.
The data $\hat X$ and $\vec p$ determine a point in the moduli space
$\calm_{g,m}$ of curves of genus $g$ with $m$ distinct marked points.
We have seen that given an $m$-tuple $\vec\lambda\eq(\lambda_1,...\,,
\lambda_m)$, we can associate to every point of $\calm_{g,m}$ a vector
space, the space $V_{\vec\lambda}(\hat X,\vec p)$ of conformal blocks.

A rather non-trivial property of these vector spaces is that they fit together
into the total space of a vector bundle over $\calm_{g,m}$.
In particular, the dimension of the spaces of conformal blocks is equal to
the rank of the vector bundle, so it depends only on the genus $g$ of the
curve (but not on the specific holomorphic structure), on the number $m$
of insertions (but not on their positions), and on the labels $\vec\lambda$.
Furthermore, the vector bundle of conformal blocks 
(sometimes called Friedan\hy Shenker bundle)
comes with additional structure, a projectively flat connection, known as
the {\em Knizhnik\hy Zamolodchikov connection\/}. It is actually a consequence 
the existence of a flat connection $\nabla\eq \rmd\,{+}\,L_{-1}\oT\rmd z$ 
on the bundles $\tilde\calh_\lambda$ over $\calm$.
The existence of this projectively flat connection immediately implies
a (projective) action of the fundamental group of $\calm_{g,m}$ on
each fiber, i.e.\ an action of the mapping class group (cf.\ \erf{1.4})
on the vector spaces of conformal blocks. (For more details and the relation 
to twisted $\cald$-modules, we refer to chapter 6 of \cite{BAki}.)

Sometimes the word ``conformal block'' is also used for (locally defined) 
horizontal sections in these vector bundles. These bundles are generically
non-trivial, i.e.\ the conformal blocks are multivalued functions of the 
insertion points. So they are not correlation {\em functions\/}; this
is the reason why above we used the words ``fields'' and ``correlation 
functions'' with quotation marks only. The condition that these sections are 
horizontal implies that they satisfy a first order differential equation.
This equation is known as the {\em Knizhnik\hy Zamolodchikov equation\/}.

For genus $g\,{=}\,1$ and one insertion of the vacuum $\Omega$, the (orbifold-)
\linebreak[0]fundamental group is ${\rm SL}(2,\zet)$ \cite{hain}; one thus 
obtains a representation of the modular group ${\rm SL}(2,\zet)$ on a complex
vector space of dimension $|I|$. In a natural basis, the generator $T$, 
acting on the complex upper half plane $H$ as $T{:}\;\tau\,{\mapsto}\,\tau{+}1$ 
of ${\rm SL}(2,\zet)$, is represented by a unitary diagonal matrix 
$T_{\lambda,\mu}$, and $S$, acting on $H$ like
  $$   S:\quad \tau\mapsto{-}1/\tau \,, $$
is represented by a unitary symmetric matrix $S_{\lambda,\mu}$. It is an 
important conjecture that the modular transformations of the characters 
discussed in Section 3.4 are the same as those of the one-point conformal 
blocks on the torus. This is the core of the {\em Verlinde conjecture}.

Up to this point, the structures we discussed are mathematically essentially
under control (see e.g.\ \cite{freN5}). There is, however, one further
crucial aspect of conformal blocks, known as {\em factorization\/}. So far we
have been talking about smooth curves. But actually, all our constructions
go through even when the curve $\hat X$ is allowed to possess certain mild
singularities, so-called ordinary double points. Such singularities can be
resolved by ``blowing up'' the double point $p$; this yields
a new curve $\hat X'$ with a projection to $\hat X$ under which $p$
has two pre-images $p'_\pm$. By factorization one means the existence
of canonical isomorphisms
  \begin{equation} g_{\hat X,\hat X'}: \quad 
  V_{\vec\lambda}(\hat X) \,\stackrel\cong\to\, \bigoplus_{\mu\in I}
  V_{\vec\lambda\cup\{\mu,\mu^+\}}(\hat X')  \labl{fact}
between the blocks on $\hat X$ and $\hat X'$.
This structure tightly links the system of bundles $\calv_{\vec\mu,g}$
over the moduli spaces $\calm_{g,m}$ for different values of $g$ and $m$.
{}From the field theoretic point of view, factorization is a novel and 
surprising structure: It relates quantum field theories defined on different 
manifolds.

Factorization can also be described from a different point of view.
One can take the curve $\hat X'$, cut out small discs around $p^+$
and $p^-$ and identify the boundaries of these discs. Taking the radius
of the disc to zero, we produce the singular curve $\hat X$. In other
words $\hat X$ can be obtained from $\hat X'$ by a {\em gluing} procedure. 
To justify factorization, frequently one invokes the physical intuition
of `inserting a complete system of states'. But to prove factorization
properties rigorously is very hard.

One consequence of factorization is that one can express the rank of 
$\calv_{\vec\lambda,g}$ for all values of $m$ and $g$ in terms of the matrix 
$S$ that we encountered in the description of the action of the modular
group. This results in the famous {\em Verlinde formula\/}, which reads
  \be  {\rm rank}\, \calv_{\vec\lambda,g} = \sum_{\mu\in I}
  |S_{\Omega,\mu}|^{2-2g}
  \prod_{i=1}^m \, \frac{S_{\lambda_i, \mu}}{S_{\Omega,\mu}} \,.  \labl3
This implies in particular that the ranks of the bundles of conformal blocks
are finite.

Using the assumption that the matrix $S$ also describes the modular 
transformations of the characters, for concrete models the matrix $S$ can be 
computed explicitly with tools from representation theory.
For WZW models, $S$ is given by the Kac\hy Peterson formula. The
combination of the Kac\hy Peterson formula for $S$ with the general
Verlinde formula (\ref3) then gives the Verlinde formula in the sense
of algebraic geometry (for reviews see \cite{beau,sorg}).

\subsection{Fusion rings}

We now use a specific type of conformal blocks, the three-point blocks on
$\complex{\dl P}^1$, to define the structure of a {\em fusion ring\/}. For every
triple of labels we introduce the non-negative integer
  $$ \caln_{\lambda_1\lambda_2\lambda_3} := \dim 
  V_{\lambda_1\lambda_2\lambda_3}(\complex{\dl P}^1) \, . $$
(A chiral CFT with the property that $\sum_{\lambda_3}\!\caln_{\lambda_1\lambda_2
\lambda_3}$ is finite for all pairs $\lambda_1,\lambda_2$ is called quasi-rational.)
Recall that there is a special label, the vacuum $\Omega$. One can show that
  \be C_{\lambda_1\lambda_2} := \caln_{\lambda_1\lambda_2\,\Omega} \labl{conj}
takes only the values 0 or 1 and is a permutation of order two; this is
called the {\em conjugation\/}. We write $\lambda^+$ for the field conjugate
to $\lambda$ and set $\caln_{\lambda_1\lambda_2}^{\,\ \ \lambda_3}\eq
\caln_{\lambda_1\lambda_2\lambda_3^+}$.

We now consider a ring (over $\zet$) with a basis $\calb\,{:=}\,
\{\Phi_\lambda\}_{\lambda\in I}^{}$, labelled by $I$ and multiplication
  \be \Phi_{\lambda_1}\,{\star}\; \Phi_{\lambda_2} =
  \sum_{\lambda_3} \caln_{\lambda_1\lambda_2}^{\,\ \ \lambda_3}\, \Phi_{\lambda_3}
  \,.  \ee
This ring is called the {\em fusion ring\/}. Factorization implies that it
is associative, commutative, semi-simple, and that $\Phi_\Omega$ acts as
the identity. A fusion ring should not be thought of as just a ring;
rather, very much like the representation ring of a Lie group,
it is a ring together with a {\em distinguished basis\/}.
As a special case of the Verlinde formula \erf3, the structure constants of the
fusion ring can be written as
  $$ \caln_{\lambda_1\lambda_2}^{\,\ \ \lambda_3} =
  \sum_{\mu} S_{\lambda_2,\mu}\, \frac{S_{\lambda_1,\mu}}{S_{\Omega,\mu}} 
  \,S_{\lambda_3,\mu}^* \,. $$
This relation can be interpreted as follows. The matrix $S$ diagonalizes
simultaneously all matrices $(\caln_{\lambda_1})_{\lambda_2}^{\;\lambda_3}
\,{:=}\,\caln_{\lambda_1\lambda_2}^{\,\ \ \lambda_3}$; this is sometimes summarized
by saying that the $S$-matrix diagonalizes the fusion rules. The eigenvalues 
of $\caln_{\lambda_1}$ are the quotients $S_{\lambda_1,\mu}/ S_{\Omega,\mu}$.
The ratio 
  $$ \cald_{\lambda} := \frac{S_{\lambda,\Omega}}{S_{\Omega,\Omega}} $$
is called the {\em quantum dimension\/} of $\lambda$. It is a real number
and satisfies $\cald\,{\ge}\,1$; for $1\,{\leq}\, \cald\,{<}\,2$, $\cald$ must 
be of the form $\cald\eq 2\cos(2\pi/n)$ for some $n\iN\zet_{\ge3}$.
(For some other aspects of fusion rings, see e.g.\ \cite{jf24}.)

The vacuum $\Phi_\Omega$ is an element of the distinguished basis that is 
invertible, with inverse again in the distinguished basis. Other labels 
$J\iN I$ with this property are of particular interest.
These so-called {\em simple currents}\,%
 \footnote{~The name simple currents is somewhat of a misnomer. While
 their fusion rules are simple, and they also correspond to simple
 superselection sectors in the sense of \cite{HAag}, they are definitely
 not currents in the sense of VOAs.}
obey
  $$ \Phi_J \star\Phi_{J^+} = \Phi_\Omega \,.  $$
Alternatively, they can be characterized by the property that the fusion
product with any element $\Phi_\lambda$ of $\calb$ 
contains just a single element, which we denote by $\Phi_{J\lambda}$:
  \be \Phi_J \star \Phi_\lambda = \Phi_{J\lambda} \, . \labl{3.5}
Yet another characterization of simple currents is that they are precisely
the fields in the distinguished basis of quantum dimension $1$.

Simple currents display a number of important properties (for a review
see \cite{scya6}).
The set of all simple currents of a rational CFT forms a finite abelian group
$\calg$, also called the {\em center\/} of a CFT. As a consequence of \erf{3.5}, 
simple currents organize the primary fields into orbits. Every simple current $J$ 
allows to assign to each label a number, the {\em monodromy charge\/} 
$Q_J(\lambda)\iN\rationals/\zet$. It is defined as the combination
  $$ Q_J(\lambda) = \Delta_J + \Delta_\lambda - \Delta_{J\lambda} \bmod \zet  $$
of conformal weights. The monodromy charge is conserved in the fusion product,
in the sense that $\caln_{\lambda_1\lambda_2\lambda_3}\,{\ne}\,0$ is
only possible if $Q_J(\lambda_1) + Q_J(\lambda_2) + Q_J(\lambda_3)\eq0 
\bmod \zet$ for all simple currents $J\iN\calg$. The elements of the matrix $S$ for 
labels on the same simple current orbit are equal up to a phase that is determined 
by the monodromy charge:
  \be S_{J\lambda,\mu} = \exp(2\pi\ii Q_J(\mu))\, S_{\lambda,\mu} \,. \labl{cent}
This relation turns out to be central for many applications of simple currents.  

In view of the applications to string theory we have in mind, we note
that every chiral CFT with at least $N\eq1$ supersymmetry possesses a non-trivial
simple current, namely the primary field $\calh_{\rm v}$ that contains the 
supercurrent $G(z)$. This simple current ${\rm v}$
has order two and conformal weight $\Delta_{\rm v}\eq3/2$; the associated monodromy 
charge takes the value $Q_{\rm v}(\lambda)\eq0$ for $\lambda$ a field in the 
Neveu\hy Schwarz sector and $Q_{\rm v}(\lambda)\eq1/2$ for $\lambda$ in the
Ramond sector. The orbits of ${\rm v}$ are just the components of a superfield. In
$N\eq2$ minimal models the simple current ${\rm v}$ is the primary field labelled as 
$\Phi^{0,2}_0$.

Superconformal chiral CFTs with $N\eq2$ supersymmetry possess in addition at 
least one more simple current, which we call $s$. It is in the Ramond sector and 
has conformal weight $\Delta_s\eq c/24$, implying that it is a Ramond ground state. 
Its order is model dependent, and its orbits are given by spectral flow. 
The monodromy charge with respect to $s$ equals half the 
U(1) charge $\bmod\,\zet$ with respect to the abelian current in the $N\eq2$
algebra. In $N\eq2$ minimal models this simple current is labelled as 
$\Phi^{0,1}_1$.

\subsection{Simple current extensions}\label{scext}

Next we introduce a concept that will be crucial for our construction
of string vacua: Simple current extensions. Let $\calg'$ be a subgroup
of simple currents of integral conformal weight $\Delta$.
Then a new VOA can be defined on the vector space
  \be  \tilde\calh_\Omega := \bigoplus_{J\in\calg'} \calh_J  \,.  \ee
We call the corresponding chiral CFT the extension of the original model by 
$\calg'$; the whole procedure is called a {\em simple current extension\/}.

On the new VOA $\tilde\calh_\Omega$ the group $G\,{:=}\,(\calg')^*$ dual 
to $\calg'$ acts by automorphisms.
For a character $\psi{:}\; \calg'\,{\to}\,\complex$ the action is defined by 
  \be  R(\psi)_{| \calh_J} := \psi(J)\, {\rm id}_{\calh_J} \, . \ee
The original VOA $\calh_\Omega$ can thus be characterized as the subalgebra 
that is left pointwise fixed under the action of $G$. Such a subalgebra
is called an {\em orbifold subalgebra\/}. Orbifolding by an abelian group
of automorphisms is the inverse
operation of simple current extension. A word of warning is in order:
Geometric orbifolds -- that is, sigma models for which the target space has the 
form of a quotient $M/G$ of a manifold $M$ -- can correspond to simple current 
extensions, to algebraic orbifolds in the sense introduced here, to a 
combination of both, or even to more general simple
current modular invariants (see Section 5.3.\ below).  

The labels and the modular matrix $\tilde S$ of the extended theory can be 
described explicitly \cite{fusS6}. To arrive at these explicit formulas 
we take for every simple current $J$ in $\calg'$ the matrix $S^J$
that describes the modular transformations of the one-point blocks on the torus
with insertion $J$. Just like $S\,{\equiv}\,S^\Omega$, for specific
models such as \wzwts\ or coset models (and thus also for $N\eq2$ minimal 
models) these matrices can be computed by representation theoretic techniques 
\cite{fusS3}. For WZW models $S^J$ coincides, up to a computable phase, with 
the ordinary S-matrix of a different affine Lie algebra, the so-called {\em 
orbit Lie algebra\/} $\breve\g$; the relevant orbit Lie algebras are given in 
the following table.\,%
 \footnote{~For a list including also the orbit \lie s for outer \auto s of
 $\g$ that are not related to simple currents, see table (2.24) of \cite{fusS3}.}
  $$ \begin{tabular}{|l|c|c|c|} \hline &&&\\[-.9em]
  \multicolumn{1}{|c|} {$\g$} &
  \multicolumn{1}{c|}  {s.c.} &
  \multicolumn{1}{c|}  {order} &
  \multicolumn{1}{c|}  {$\breve\g$}
  \\[1.2mm] \hline\hline &&&\\[-2.8mm]
   $A\untw_n$      & $\!\!J^{(n+1)/N}\!\!$ & $\!N\!<\!n\!+\!1\!$
                   & {$\!A\untw_{((n+1)/N)-1}\!$} \\[1.9mm]
   $A\untw_n$      & \sicu  &$n+1$&$ \{0\} $                \\[1.9mm]
   $B\untw_{n}$    & \sicu  & 2  & $ \tilde B_{n-1}\twtw $  \\[1.9mm]
   $C\untw_{2n}$   & \sicu  & 2  & $ \tilde B_n\twtw  $     \\[1.9mm]
   $C\untw_2$      & \sicu  & 2  & $ A\twtw_1 $             \\[1.9mm]
   $C\untw_{2n+1}$ & \sicu  & 2  & $ C\untw_n $             \\[1.9mm]
   $D\untw_n$      & \jv    & 2  & $ C\untw_{n-2}$          \\[1.9mm]
   $D\untw_{2n}$   & \js    & 2  & $ B\untw_n $             \\[1.9mm]
   $D\untw_{2n+1}$ & \js    & 4  & $ C\untw_{n-1}$          \\[1.9mm]
   $E\untw_6$      & \sicu  & 3  & $ G\untw_2 $             \\[1.9mm]
   $E\untw_7$      & \sicu  & 2  & $ F\untw_4 $            
   \\[.4em] \hline \end{tabular} 
  $$
Here the following notation is used.
$\g\eq X_n\untw$ is the untwisted affine \lie\ with horizontal subalgebra
$X_n$; $\{0\} $ is the trivial zero-dimensional \lie. Only one series of
twisted affine Lie algebras, denoted by $\tilde B_n\twtw$, appears:
It is the only such series for which the characters of integrable \rep s 
span a module of ${\rm SL}(2,\zet)$. Whenever the simple current
is unique up to conjugation, we have denoted it by $J$; conjugated simple
currents lead to identical orbit \lie s. For $D_n$, there is both the vector
simple current \jv, of order 2 and conformal weight $k/2$, and two
spinor simple current \js\ of conformal weight $kn/8$. For $A_n$ there are 
$n{+}1$ simple currents, which can be written as powers of one of them, which
has order $n{+}1$ and which we denote by $J$.

The simple current symmetry \erf{cent} generalizes to
  \be  S^J_{K\lambda,\mu} = F_\lambda(K,J) \exp(2\pi\ii Q_K(\mu))\,
  S^J_{\lambda\mu} \,. \labl{FQ} 
The action of $\calg'$ is typically {\em not\/} free. Accordingly one 
associates to every label $\lambda$ its {\em stabilizer subgroup}
  $$ \cals_\lambda := \{ J\iN\calg' \,|\, \Phi_J{\star}\,\Phi_\lambda\eq\Phi_\lambda\} 
  \,. $$
A label with non-trivial stabilizer is also called a {\em fixed point\/}.
$S^J_{\lambda,\mu}$ can be non-zero only if both $\lambda$ and $\mu$
are fixed points of $J$. The quantities $F_\lambda$ appearing in formula 
\erf{FQ} can be used to define the subgroup 
  \be \calu_\lambda := \{ J\iN\cals_\lambda
  \,|\, F_\lambda(K,J)\eq1\;{\rm for\;all}\; K\iN\cals_\lambda\}  \labl{ustab}
of the stabilizer $\cals_\lambda$, the so-called {\em untwisted stabilizer\/}.
The quantity $F_\lambda$ has a cohomological interpretation \cite{bant6}, 
which implies in particular that
  $$ d_\lambda := \sqrt{{|\cals_\lambda|}\,/\,{|\calu_\lambda|}} $$
is a non-negative integer.

The labels of the extension are then equivalence classes of pairs
$(\rho,\psi)$, where $\rho$ is a label with $Q_J(\rho)\eq0$ for all
$J\iN\calg'$ and $\psi$ is a character on the untwisted stabilizer \erf{ustab}.
Two labels $(\rho,\psi)$ and $(\rho',\psi')$ are considered to be equivalent if 
there is a simple current $J\iN\calg'$ such that $\rho'\eq J\rho$ and
$\psi'(K)\eq F_\rho(J,K) \psi(K)$. The fact that a character
$\psi$ must be introduced to label fields with non-trivial untwisted stabilizer
is known as {\em fixed point resolution\/}. We emphasize that
only labels of vanishing monodromy charge $Q_J(\rho)\eq0$ appear. This 
property allows us to implement projections in string theory.

The ease with which projections can be implemented with the help of simple 
currents has also been used in applications of chiral CFT to the fractional
quantum Hall effect \cite{fpsw}. In that context the electron corresponds to a
simple current with half-integer conformal weight and the constraints stem
from the physical locality requirement of Fermi statistics of the electron.
(Note, though, that in this application one is not dealing with an extension.)

The characters of these fields are given, as a function of $\tau\iN H$, by
  \be  \chii_{[\rho,\psi]}(\tau) = d_\rho \sum_{J\in\calg'/\cals_\rho}
  \chii_{J\rho}(\tau) \,.  \Labl xc
This determines also the modules of the extension, at least when regarded
as modules over the original chiral algebra. Finally, the modular matrix 
$\tilde S$ of the extension reads
  \be   \tilde S_{[\lambda\psi],[\lambda'\psi']}
  = \frac{|\calg'|}{\sqrt{|\cals_\lambda| |\calu_\lambda|}
  \sqrt{|\cals_{\lambda'}| |\calu_{\lambda'}|}}
  \sum_{J\in\calu_\lambda\cap\calu_{\lambda'}}\psi(J)\, S^J_{\lambda,\lambda'}
  \,\psi'(J)^* \, . \ee

We conclude with a first application of simple extensions that will
be quite important for the construction of string vacua. As we have seen,
the ordinary tensor product $\calc_1{\times} \calc_2$ of superconformal 
theories is not supersymmetric. We can, however, consider a simple
current extension of this tensor product, namely by the tensor product
of simple currents $({\rm v}_{(1)},{\rm v}_{(2)})$. This simple current
has integral conformal weight and hence can indeed serve as an extension.
In the extended theory, the two fields $G(z)\oT\bfe$ and $\bfe\oT G(z)$
belong to the same module, and it makes sense to consider their sum; this
provides a supercurrent for the extended theory. Another effect of the extension 
is that only fields $(\lambda_1,\lambda_2)$ with vanishing monodromy charge 
survive. This is equivalent to the statement that the two fields are either both
in the Neveu\hy Schwarz or both in the Ramond sector. Thus the extension
aligns the periodicity of fermionic fields on the world sheet.
For more than two, say $r$, factors in a tensor product, we must
extend by the $(\zet_2)^{r-1}$ group of simple currents of the form
$({\rm v}_{(1)}^{\eps_1},{\rm v}_{(2)}^{\eps_2},...\,,{\rm v}_{(r)}^{\eps_r})$,
where $\eps_i$ takes the values $0$ or $1$ and $\sum_i \eps_i$ is even.

This prescription can also be understood in the language of super VOAs.
These are VOAs equipped with an additional grading over $\zet_2$, 
$\calh_\Omega\eq\calh_\Omega^{(\bar 0)}\,{\oplus}\,\calh_\Omega^{(\bar 1)}$,
with the requirement that field-state correspondence preserves the parity:
$Y(v,z)$ is a formal series of even operators if $v$ is in the bosonic
subalgebra, $v\iN\calh_\Omega^{(\bar 0)}$, and of odd operators if $v$ is
in the fermionic part, $v\iN\calh_\Omega^{(\bar1)}$. One requires that bosonic
states have integral conformal weight and that fermionic states have conformal 
weight in $\frac12\,{+}\,\zet$. Notice that the bosonic subalgebra is a VOA;
this is the VOA we have been considering for supersymmetric theories. 
$\calh_\Omega^{(\bar 1)}$ is a module over this VOA, and for supersymmetric 
theories this is just the module $\calh_{{\rm v}}$. The tensor product
of two super VOAs has 
  $$\left(\calh_\Omega^{(\bar 0;1)}{\otimes} \calh_\Omega^{(\bar 0;2)}\right) 
  \oplus \left(\calh_\Omega^{(\bar 1;1)}{\otimes} \calh_\Omega^{(\bar 1;2)}\right)
  $$ 
as its bosonic part. This provides yet another explanation for the presence 
of the simple current extension by 
$\calh_{{\rm v}}^{(1)}{\otimes}\calh_{{\rm v}}^{(2)}$ in the 
supersymmetric tensor product.

To summarize: For supersymmetric theories a tensor product should always 
be accompanied by an appropriate simple current extension.  

\section{Full CFT}

We are now finally in a position to address full CFT. The most important
point of this section is that the construction of a full CFT from a chiral
CFT can be formulated in completely model-independent terms. In this
transition chiral data related to the action of the mapping class group
on the spaces of conformal blocks -- S-matrices, fusing matrices, 
conformal weights and the like -- enter.

\subsection{The central task}

It should have become apparent that chiral CFT 
is not a local quantum field theory in the usual sense and that conformal 
blocks, being multivalued, are not genuine correlation functions.
Moreover, we have defined the theory on a complex curve, but for
applications we wish to set up full CFT on a \twodim\ conformal manifold $X$.

Suppose we are interested in correlators on $X$ of a full CFT that is
based on some given chiral CFT. 
We want to relate them to conformal blocks on the double $\hat X$ of $X$.
To this end, we must lift the situation from $X$ to $\hat X$.
Boundary points on $X$ have a unique pre-image on $\hat X$. Thus, in order to
define conformal blocks, we need to attach one chiral label to each
boundary insertion. A bulk point, in contrast, has two pre-images on
$\hat X$, and hence requires two chiral labels.
    
In the case of the sphere, $X\eq S^2$, the cover $\hat X$ consists of two 
spheres with opposite orientation. As quasi-global complex coordinates on 
the two components of $\hat X$, we choose $z$ and $\tilde z$. Then the 
involution $\sigma$ maps the point with coordinate $z$ to a point on the other 
component with complex conjugate coordinate: $\tilde z\eq z^*$. For this 
reason, the coordinate $\tilde z$ is more commonly, and more suggestively, 
denoted by $\bar z$. But one should be aware of the fact that in chiral CFT 
$\tilde z$ and $z$ are indeed two completely independent complex variables.
A frequent description in the literature is to say that ``one starts with
a single complex variable $z$ and its complex conjugate and then treats the 
two variables as formally independent''.  When it comes to concrete 
calculations this statement amounts to nothing else than working with the 
Schottky double to treat the chiral aspects of CFT.

The observations about bulk and boundary insertions supply us with
enough structure to define conformal blocks for the
double, and we get a vector bundle $\calv$ over the moduli space 
$\calm_{\hat X}$ of the double. (We will always have in mind surfaces with 
marked points and associated insertion labels, but for brevity suppress 
them in our notation.) There is, however, no embedding of the 
moduli space $\calm_X$ of $X$ itself into $\calm_{\hat X}$; only the 
Teichm\"uller space $\calt_X$ of $X$ can be embedded into the Teichm\"uller
space $\calt_{\hat X}$ of the double. 

The construction is therefore slightly more involved: Using the fact
that the Teichm\"uller space is a covering of the moduli space, we 
first pull back the bundle $\calv$ by the covering map to $\calt_{\hat X}$
and then restrict it to the submanifold $\calt_{X}$. 

We have now obtained a vector bundle $\calv$ of conformal blocks over the 
Teichm\"uller space $\calt_{X}$. The central task of full CFT can then
be phrased as follows:
  \begin{quote}
  {\sl Select a (global) horizontal section of this bundle in such a way\\
  that it is just the correlation function we are looking for.}
  \end{quote}

In fact, what we want to obtain is not just the correlation function
as a function on Teichm\"uller space. Rather, we want to know the
correlator as a function of the moduli of $X$ and of the insertion points, 
i.e.\ we are looking for a function on the {\em moduli\/} space $\calm_X$.\,
For non-vanishing Virasoro central charge, this is actually too strong a 
requirement, due to the {\em Weyl anomaly\/}. In string theory, however, the 
total central charge vanishes; we therefore suppress this subtlety in the 
present review. Moreover, since in string theory one wishes to integrate CFT 
amplitudes over the moduli space, one needs a volume form over the moduli 
space $\calm_X$; indeed, a natural measure is provided by the ghost system.

Recall from Section 2.4 that the moduli space $\calm_X$ can be
obtained from the Teichm\"uller space $\calt_X$ by modding out
the relative modular group $\Omega_\sigma(\hat X)$.
Correlation functions should be genuine functions on $\calm_X$, because
this ensures that they are functions of moduli and insertion points.
This holds for those sections in the restriction of $\calv$ to
$\calt_X$ that are invariant under the action of the relative modular group.
Modular invariance is therefore a locality requirement. It is the first
condition we impose on correlation functions. But the
requirement of locality is too weak to fix them uniquely.
One must in addition impose compatibility with the factorization properties
(or equivalently, the gluing properties) \erf{fact} of the conformal blocks. 

There are two types of factorization constraints, corresponding to bulk and 
boundary insertions, respectively. First we study what happens when we glue 
together two bulk insertions on a world sheet $X$. On
the double, this amounts to a simultaneous gluing of two pairs of
insertions. For the correlation functions $C(X)$, one thus requires
  $$ C(X) = \sum_{\lambda\in I} c_{\lambda}^{}\, (g_{\hat X,\hat X'})^{-1}\,
  C(X'_\lambda) \,, $$
where $c_\lambda$ is the normalization of the bulk two-point function
and $g_{\hat X',\hat X}$ is the isomorphism \erf{fact}.
It is important to notice that $c_{\lambda}$ is non-zero only for those
bulk fields that appear in the modular invariant partition function. In this
sense the factorization constraint guarantees that only physical fields
propagate in `loop channels'.

Similarly, one can glue two boundary insertions. In this case one
deals with a single gluing on the double. One finds
  $$ C(X') = \sum_{\lambda\in I} \tilde c_\lambda^{} \,
  (g_{\hat X',\hat X})^{-1} \, C(X'_\lambda) \,, $$
where $\tilde c_\lambda$ is the normalization of the two-point functions
of boundary fields. 
At this point, it is far from obvious whether there exists any solution to
these constraints, nor whether possible solutions are unique. In fact,
only little is known about uniqueness whereas, at least in the simplest cases,
the existence of a consistent solution has been established, see below.

We conclude this subsection with two possible generalizations of the
construction. The first one concerns the case when we consider full CFT
only on closed and orientable surfaces $X$. Then the double 
has two disjoint components, corresponding to left movers and right movers. 
This allows us to consistently use two {\em different\/} chiral conformal 
field theories on the two components. The models obtained this way
are called {\em heterotic\/}; the heterotic string is the most prominent
example. Other examples are asymmetric orbifolds. We will see below how
simple currents allow to construct also heterotic theories.
When one considers non-heterotic CFT on closed oriented surfaces only, the 
central task has also been formulated as follows: Find hermitian scalar
products on all bundles of conformal blocks that are compatible with 
factorization. These scalar products provide us with
genuine functions, the correlation functions.

The second generalization concerns the other extreme: When we allow also for
surfaces $X$ with boundaries, we may decide to use on the 
double $\hat X$ of a surface $X$ with non-empty boundary not the conformal
blocks for the original chiral algebra, but those for a subalgebra of
the chiral algebra. Since the chiral algebra encodes the chiral symmetries,
such a setup allows us to describe boundary conditions
that preserve only part of the bulk symmetries.  

\subsection{Modular invariance}

As an example we consider the torus $X\eq T^2$ with a single insertion 
of the vacuum. Its double consists of two disjoint copies of a torus, with 
opposite orientation: $\hat X\eq (+T^2)\,{\cup}\,(-T^2)$. The Teichm\"uller 
space of the torus is the complex upper half plane, so the mapping 
class group of the double is the product of two copies of ${\rm SL}(2,\zet)$. 
The double $\hat X$ is characterized by the values $(\tau,-\tau^*)$ of the 
modular parameters of the toroidal connected components.  This can be seen as 
follows: The anticonformal map on the plane from which the torus is obtained by 
dividing out a lattice is complex conjugation. It maps a lattice 
with fundamental cell $0,1,\tau$ and $\tau{+}1$ to a fundamental cell 
$0,1,\tau^*,\tau^*{+}1$, which is equivalent by reflection about the origin to 
a torus with parameter $-\tau^*$. As a consequence, the relative modular group 
is the diagonal subgroup of ${\rm SL}(2,\zet)\,{\times}\, {\rm SL}(2,\zet)$ 
acting by $(S,S^*)$ and $(T,T^*)$ on the product of the two upper half planes. 
Once one identifies characters and one-point blocks, invariance under the 
relative modular group implies ordinary modular invariance
  \be [Z,S] = 0 = [Z,T]   \labl{mod} 
for the partition function
  $$ Z(\tau) = \sum_{\lambda,\tilde \lambda\in I} Z_{\lambda\tilde \lambda}\, 
  \chii^{}_{\lambda}(\tau)\, \chii_{\tilde\lambda}(\tau)^* \,.  $$ 
Here we used that
  $$ (\chi_\lambda(\tau))^*_{} = \sum_{n\ge0} d_n\, 
  \eE^{2\pi(-\ii)\tau^*(\Delta+n-c/24)} = \chi_\lambda(-\tau^*_{}) \, . $$

The constraints \erf{mod} always admit two canonical solutions: The identity 
$Z_{\lambda\mu}\eq\delta_{\lambda\mu}$, and
the conjugation matrix \erf{conj}, $Z_{\lambda\mu}\eq C_{\lambda\mu}$.
More generally, together with any $Z_{\lambda\mu}$, 
$(ZC)_{\lambda\mu}^{}\eq Z_{\lambda\mu^+}$ is a modular invariant as well.

Modular invariance is a very strong requirement on the completeness of
the theory. In {\em closed\/} string theory, it implies that all anomalies
of the corresponding space-time theory vanish \cite{Lesw}. The converse is 
not true (this is not always appreciated); it is e.g.\ easy
to construct anomaly free spectra that violate modular invariance.
For us the fact that a consistency condition on the \twodim\ world sheet theory
is more restrictive for the space-time theory than ordinary field-theoretic
consistency conditions is one of the most amazing features
of string theory.

\subsection{Simple current modular invariants}

Simple currents allow to set up a remarkably simple algorithm \cite{krSc} for
constructing solutions to the modular invariance condition \erf{mod}.\,%
 \footnote{~There are also exceptional modular invariants that cannot be 
 obtained this way. This includes e.g.\ certain invariants stemming from
 so-called conformal embeddings, compare e.g.\ \cite{boev5,zube8}.}
These solutions will enter our discussion of string theory in two different 
ways: They serve to implement projections in string theory, and they allow 
to construct the mirror of a Gepner compactification. Let us describe the 
construction of simple current modular invariants in some detail:\\[.2em]
1)~Recall that all simple currents form a finite abelian group, the
   center $\calg$. The {\em effective center} is defined as the subgroup
   of $\calg$ consisting of those simple currents $J$ whose conformal
   weight $\Delta_J$ multiplied by the order $N_J$ of $J$ is integral.
   The first step is to choose a subgroup $\calg'$ of the effective center. \\[.2em]
2)~Like any finite abelian group, $\calg'$ can be written as a product of 
   cyclic groups,
  $$ \calg' = \zet_{N_1} \times \cdots \times \zet_{N_k} \, . $$
   We choose generators $J_i$ of order $N_i$ for the cyclic groups $\zet_{N_i}$.
   \\[.2em]
3)~Compute the relative monodromy charges:
  $$ R_{ik} := Q_{J_i}(J_k) = Q_{J_k}(J_i) \bmod \zet \, . $$
The diagonal elements of the matrix $R\eq(R_{ik})$ must be defined  even modulo 
$2\zet$. Indeed, one can show that the conformal weight $\Delta(J_k)$ obeys
  \be \Delta(J_k) = (N_k\,{-}\,1) \frac{R_{kk}}2 \bmod \Frac12\zet \, . 
  \labl{rdef}
We define $R_{kk}$ modulo $2\zet$ such that the identity \erf{rdef} holds
modulo $\zet$. \\[.2em]
4)~Choose a matrix $X$ subject to the two conditions
  \be X + X^{\rm t} = R\,,
  \qquad X_{ij} \in \frac{\zet}{{\rm gcd}(N_i,N_j)} \bmod\zet \,.  \labl{dt}
Thus the symmetric part of $X$ is fixed by the relative monodromies,
while the antisymmetric part is free. 
The possibilities are in fact in one-to-one correspondence to the elements of
the cohomology group $H^2(\calg',\complex^{\times\!})$. By analogy with geometric 
orbifolds, the antisymmetric part of $X$ has been called {\em discrete torsion\/} 
\cite{krSc}. Recall, however, that the relationship between 
geometric orbifolds and simple current modular invariants is rather subtle
in general. Analogous subtleties appear in the connection between the
antisymmetric part of $X$ and discrete torsion in geometric orbifolds.
Still, these considerations illustrate that discrete torsion is a structure that 
appears naturally in CFT; it is neither `inherently stringy', nor does it require 
a geometric interpretation of the CFT. When a sigma model formulation of a CFT
is available, then the presence of discrete torsion amounts to having a
non-trivial action of the orbifold group on the antisymmetric tensor field
\cite{shaR6}.

\smallskip
Given all these data, one can show that the matrix $Z$ whose only non-zero 
entries read
  \be Z_{\lambda^+_{\phantom i}\!,J\lambda} = |\cals_\lambda|\,
  \prod_{i=1}^k \delta^{(\zet)}(Q_{J_i}(\lambda) \,{+}\, \mbox{\large$\sum_j$}
  X_{ij}\beta_j) \labl{krSc}
are solutions to \erf{mod}. Here $\delta^{(\zet)}(x)$, defined for real arguments, 
takes the value 1 for integral $x$ and is zero else, and we expressed 
$J$ as $J\eq \prod_{j=1}^k\! J_j^{\beta_j}$ through the generators $J_i$.

According to quite general arguments \cite{Mose}, a modular invariant is of the 
following form.  One must choose an extension ${\mathfrak A}_L$ of the chiral 
algebra for left movers and a possibly different extension ${\mathfrak A}_R$ 
for the right movers. We denote the characters of these extensions by 
$\chii_{\lambda,L}$ and $\chii_{\tilde\lambda,R}$, respectively. The two 
extensions must have the property that the two fusion rings are isomorphic, 
and more specifically, that there is an isomorphism $\omega$ such that for 
each $\lambda$ the difference between the conformal weights of $(\lambda,L)$
and $\omega(\lambda,L)$ is integral. Every modular invariant is then of the form
  $$ Z(\tau) = \sum_{(\lambda,L)} \chii_{(\lambda,L)}^{}\,
  \chii_{\omega(\lambda,L)}^{} \, . $$

For the modular invariants \erf{krSc}, both ${\mathfrak A}_L$ and
${\mathfrak A}_R$ are simple current extensions. The simple currents appearing
in ${\mathfrak A}_R$ are those in the right kernel and the ones in
${\mathfrak A}_L$ are those in the left kernel of the matrix $X$. For a pure
extension (all $J\iN\calg'$ of integral conformal weight and $X\eq 0$),
${\mathfrak A}_R$ and ${\mathfrak A}_L$ coincide, and the prescription 
\erf{krSc} reduces to
  $$   Z(\tau) = \!\!\!\sum_{\scriptstyle [\mu]:\ \  \mu\in I,\ \ \atop
  \scriptstyle Q_J(\mu)=0 \forall J\in\calg'}\!\!\!\!\! \mbox{\Large(}
  \, |\calu_\mu|\cdot \mbox{\Large$|$} d_\mu \sum_{J\in\calg/\cals_\mu}\!
  \chii_{J\mu}(\tau) \mbox{\Large$|$}^2 \mbox{\Large)}
  = \!\!\sum_{\scriptstyle [\mu]:\ \  \mu\in I,\ \ \atop
  \scriptstyle Q_J(\mu)=0 \forall J\in\calg'}\!\!\! \sum_{\psi\in\calu_\mu^*}
  \mbox{\large$|$} \chii_{[\mu,\psi]}(\tau) \mbox{\large$|$}^2 \,,  $$
where $\calu_\mu$ is the untwisted stabilizer \erf{ustab} and in the second
equality the formula \Erf xc for the extended characters
$\chii_{[\mu,\psi]}(\tau)$ is used.

\subsection{Boundary conditions and four-point blocks on $\complex{\dl P}^1$}

We now turn to solutions of the factorization constraints including also
surfaces with boundaries. As it turns out, even after fixing a modular 
invariant torus partition function, the solution to these constraints is in 
general not unique. It is generally expected that the set of solutions
can be parametrized by assigning to every boundary segment a {\em boundary 
condition\/}. When referring to segments of the boundary, also any insertion 
point on the fixed point set of 
$\sigma$ must be regarded as separating the boundary into different segments. 
One must assign boundary conditions to each of these boundary segments, and 
to each insertion point one must assign a {\em boundary field\/}. Such a boundary
field $\Psi$ carries, in addition to the chiral label $\mu$, the two 
labels $a,b$ of the adjacent boundary conditions, as well as a degeneracy 
label $A$ that distinguishes among inequivalent ways of transforming one of 
these boundary conditions into the other. The full labelling of a boundary
field is therefore $\pso\mu abA$.
In particular, for $a\ne b$ a boundary field plays the role of changing the
boundary condition; on the other hand, boundary fields are not taking part
in the characterization of the boundary conditions themselves. We can therefore 
regard a boundary condition as a solution to the factorization constraints 
for surfaces $X$ with a single boundary component and only bulk insertions. 
Moreover, factorization (e.g.\ of the M\"obius strip to a crosscap plus a 
disc) allows us to restrict our attention to the case where $X$ is the disc
and where there is a single bulk insertion.

The situation when the bulk partition function $Z$ is given by conjugation and
when the boundary conditions preserve all bulk symmetries (so that one and the 
same chiral algebra can be used to construct conformal blocks on the covers of 
all surfaces) has been called the
{\em Cardy case\/}. Remarkably, in that case all correlators for any topology of
$X$ can be described using techniques from topological field theory.
In that framework, factorization properties and invariance under the
relative modular group can be rigorously proven. (A short review of this
approach can be found in \cite{scfu3}.)

Thus, to determine individual boundary conditions we should study 
correlators of bulk fields on a disc. At the chiral level, these 
correlators correspond to conformal blocks on the sphere $\complex{\dl P}^1$, 
with an even number of insertions. 
The moduli space of three or less points on a sphere is trivial (see 
below), so non-trivial constraints arise for the first time from the
four-point blocks. These blocks appear also in the familiar case of
correlators of four bulk insertions on $X\eq S^2$, as well as for
four boundary insertions on the disc. The former case is treated in standard 
texts on CFT, and we will not discuss it any further, while the latter 
situation is described in some detail below.

To examine boundary conditions on the disc, we introduce the bulk-boundary 
operator product
   \be  \Phi_{\mu\mu^+}(z) \ {\sim}\ \sum^{}_{\nu\in I} \sum^{}_B\,
   (1{-}|z|^2)^{-2\Delta_\mu+\Delta_\nu}_{}\, {\calr}^a_{\mu\mu^+\nu}
   \pso\nu aaB({\rm arg}\,z) \quad\ \mbox{for\,\ }|z|\,{\to}\,1 \,.  \labl16
This tells us what happens when a bulk field $\Phi_{\mu\mu^+}$ approaches
the boundary of the disc $|z|\,{\le}\,1$ with boundary condition $a$: It creates
excitations on the boundary, described by the boundary operators $\pso\nu aaB$.

Two factorizations of this amplitude are possible. One may first use
the bulk OPE to produce a single bulk field and then consider its 
bulk-boundary operator product; the resulting expression contains the
{\em reflection coefficient\/} $\calr^a$ once. The other factorization is to 
apply \erf{16} to both bulk fields; then two reflection coefficients
$\calr^a$ appear. Comparison of the two factorizations and specialization
to the case when the chiral label $\nu$ of the boundary field is the vacuum 
$\Omega$ yields an identity of the form
  \be  
  \calr^a_{\lambda_1\lambda_1^+\Omega} \, \calr^a_{\lambda_2\lambda_2^+\Omega}
  = \sum_{\lambda_3\in I} \tilde\caln_{\lambda_1\lambda_2}^{\,\ \ \lambda_3}\,
  \calr^a_{\lambda_3\lambda_3^+\Omega} \,, \labl{class}
where  $\tilde\caln_{\lambda_1\lambda_2}^{\,\ \ \lambda_3}$ is some complicated
combination of operator product coefficients and representation matrices for
$\pi_1(\calm_{4,0})$ acting on four-point blocks.

One {\em structural\/} information can immediately be read off
\erf{class}: For fixed boundary condition $a$ the quantities $\calr^a$ form,
a \onedim\ \rep\ of an associative algebra with structure constants 
$\tilde\caln_{\lambda_1\lambda_2}^{\,\ \ \lambda_3}$, this algebra is called
the {\em classifying algebra\/}. The problem of classifying boundary conditions
is thereby reduced to the problem of studying the representation theory of 
the classifying algebra.

In the Cardy case, it is found that the structure constants 
$\tilde\caln_{\lambda_1\lambda_2}^{\ \ \lambda_3}$ are just the fusion rules. 
The \onedim\ representations of the fusion algebra are the (generalized) quantum
dimensions. Indeed, it follows from the Verlinde formula \erf3 that 
for each $a\iN I$ there is one \irrep\ $R_a$, reading
  $$ R_a(\Phi_\lambda) = \frac{S_{\lambda,a}}{S_{\Omega,a}}  \, . $$
It follows in particular that the boundary conditions in the Cardy case are in
one-to-one correspondence to the primary fields.

The reflection coefficients $\calr$ also determine the correlation functions of
a single bulk field on a disc with boundary condition $a$. The information
on these correlators is frequently summarized in a single quantity, the 
so-called {\em boundary state\/} $|a\rangle\!\rangle$. Just like the Ishibashi 
states \erf{ishi}, this is not an element of the space of bulk states, but a 
bilinear form on that space. The correlator for the bulk field 
$\Phi(v\ot\tilde v;z{=}0)$
inserted in the center of the disc $|z|\,{\le}\,1$ is given
by the value of $|a\rangle\!\rangle$ on $v\oT\tilde v$:
  $$ {\langle\Phi(v\ot\tilde v;z{=}0)\rangle}_a =
  \langle\!\langle a\, |\, v\oT \tilde v\rangle \, . $$
(Other positions of the bulk field can be related to this
case using the action of the M\"obius group ${\rm SL}(2,\reals)$ on the disc.)
In the Cardy case, the boundary state is a linear combination of
Ishibashi states. With a suitable normalization of bulk fields, it reads
  \be |a\rangle\!\rangle = \sum_{\lambda\in I} 
  \frac{S_{\lambda,a}}{\sqrt{S_{\Omega,\lambda}}}\, |B_\lambda\rangle \, . 
  \labl{bs}

This analysis has been generalized beyond the Cardy case \cite{fuSc1112}.
We do not describe those results here, but indicate the main idea:
The fusion rules, which are the structure constants of the classifying algebra
in the Cardy case, are the dimensions of the spaces of three-point blocks.
A dimension is the trace of the identity operator. Similarly, in the more 
general situation, the structure constants of the classifying algebra can be 
obtained as traces of non-trivial operators $\Theta$ on the conformal blocks:
  $$ \tilde\caln_{\lambda_1\lambda_2\lambda_3} 
  = {\rm tr}^{}_{V_{\lambda_1\lambda_2\lambda_3}} \Theta \, . $$
For details, we refer to a recent review \cite{fuSc13}. We also remark that 
the one-point functions on the real projective plane $\reals{\dl P}^2$, the 
crosscap, are constrained by factorization constraints on four-point blocks as 
well; for a review of the so-called cross-cap constraint we refer to \cite{prad}.

Factorization arguments allow to compute the zero-point amplitude of an
annulus with boundary conditions $a$ and $b$ from the boundary state:
  $$ A_{ab}(t) = \langle a|\, \eE^{-(\pi\ii/t)\,
  (L_0\ots\bfe+\bfe\ots L_0-c/12)} \,|b\rangle \, . $$
In the Cardy case, this leads to the result
  \be A_{ab}(t) = \sum_{\nu\in I} \caln_{ab}^{\;\ \nu}\,
  \chii_\nu (\Frac{\ii t}2) \,, \labl{cardy}
which tells us that the so-called annulus multiplicities are equal to the 
fusion rules.

Now recall that in the case of the torus, we could regard the zero-point 
amplitude as a partition function, counting closed string states. The annulus
is a one-loop diagram, too, and we will interpret it as the partition
function for open string states. Under (non-chiral) field-state 
correspondence, these states give rise to {\em boundary fields}. 
The partition function \erf{cardy} tells us in particular what the precise
meaning of the  degeneracy label $A$ of the boundary fields $\pso\lambda abA$
is in the Cardy case: It specifies the chiral coupling between $\lambda$, $a$
and $b$ and hence can take the values $A\eq1,2,...\,,\caln_{ab}^{\ \ \lambda}$.

Just like the bulk fields, the boundary fields should satisfy an operator 
product expansion, which schematically reads
  \be  \pso\lambda abA(x)\,\pso\mu bcB(y)
  \,\sim\, \sum_\nu \sumn L\lambda\mu\nu \suma C ca\nu
  \tC{\lambda\mu}\nu abcLABC \, [\pso \nu acC(y) + ...\,] \,,  \Labl tC
where ${\rm L}\iN\{1,2,...\,,\N\lambda\mu\nu\}$ labels a basis of the space
of chiral couplings from $\lambda$ and $\mu$ to $\nu$. 
The constants $\tC{\lambda\mu}\nu abcLABC$ can be determined by a similar
method as the one that yields the OPE of bulk fields. The moduli space of 
four points on $\complex{\dl P}^1$ is one-dimensional: The invariance of
$\complex{\dl P}^1$ under the M\"obius group ${\rm SL}(2,\complex\,)$ allows to 
fix the positions of three points, say to $z_1\eq0,\,z_2\eq1$ and 
$z_3\eq\infty$. The position $z$ of the fourth point is the only remaining 
coordinate. The moduli space $\calm_{0,4}$ has singularities for 
$z\,{\to}\,z_1,z_2$ and $z_3$; one calls these three limits the $s$-, $t$-, 
and $u$-channel, respectively. In these limits $z\,{\to}\,z_i$, the conformal 
blocks possess natural expansions, which can be compared with the help of 
parallel transport with respect to the Knizhnik\hy Zamolodchikov
connection. The comparison of the four-point correlators in the different 
channels yields restrictions both for four bulk and for four boundary fields.
Notice that in both cases the four-point blocks on the sphere are again
the crucial ingredient to determine OPEs.

In the Cardy case, it can be shown that for every (rational) \cft, the 
structure constants $\tC{\lambda\mu}\nu abcLABC$ are nothing but suitable 
entries of fusing matrices {\sf F}: 
  \be  \tC{\lambda\mu}\nu abcLABC
  = \mbox{\large(} \F{\sxs L\nu C}{\sxs A{b^+_{\phantom i}}B}{a^+_{\phantom i}}
  {\lambda\;}{\;\mu_{}} c {\mbox{\large)}}^* \,. \Labl1f
(Fusing matrices relate the four-point blocks on the sphere in the s-channel 
to those in the t-channel.) The proof relies on the pentagon equation and the 
tetrahedral symmetry for {\sf F}. The relation \Erf1f can indeed be 
generalized beyond the Cardy case: The structure constants of the boundary 
OPE are fusing matrices of more general fusion rings, see \cite{scfu3}. (For a 
different approach to boundary OPEs beyond the Cardy case, see \cite{zube8}.)

\subsection{Boundary conditions in a general simple current modular invariant}
\label{bsc}

We now study the boundary conditions for an arbitrary simple
current modular invariant. This will be a crucial input in our discussion
of D-branes in Gepner models. We consider a full CFT with
the modular invariant given by \erf{krSc} and wish to describe all 
boundary conditions for the extended theory that preserve (at least)
the original chiral algebra ${\mathfrak A}$.\,%
 \footnote{~This construction covers huge class of boundary conditions, but in
 general it does {\em not\/} exhaust the space of all conformal boundary
 conditions of a CFT. In particular there exist boundary conditions without
 automorphism type; known examples include boundary conditions associated to
 conformal embeddings (for a review see \cite{zube8,scfu3} and lectures by
 J.-B.\ Zuber in this volume) and certain boundary conditions
 for the $\zet_2$-orbifold of a compactified free boson \cite{fuSc1112}.}

We wish to generalize \erf{bs}. To this end we must give three types of data:
The Ishibashi states, the labels $a$ for the boundary conditions, and
a generalization of the quotient of S-matrix elements. We will see that
the latter are again chiral data associated to the original chiral algebra.

The following prescription \cite{fhssw} gives the correct
result. The Ishibashi states are the two-point blocks
on the sphere for the ${\mathfrak A}$-theory. They correspond to terms 
$Z_{\lambda\lambda^+}$
in the partition function \erf{krSc}, and they are given by {\em pairs\/}
$(\lambda,J)$ with $J\iN\cals_\lambda$ and $Q_K(\lambda)\,{+}\,X(K,J)\iN\zet$
for all $K\iN\calg'$.

The boundary conditions take values in a different set. To define
it, we must generalize the untwisted stabilizer
\erf{ustab} to the central stabilizer $\calc_\lambda$:
  \be  \calc_\lambda :=\{ J\iN\cals_\lambda \,|\, \phi_\lambda(K,J)
  \eq1 \mbox{ for all } K{\in}\,\calg' \} \ee
with
  $$  \phi_\lambda(K,J) := \eE^{2\pi\ii X(K,J)}\, F_\lambda(K,J)^* \,.  $$
This is again a subgroup of the stabilizer $\cals_\lambda$. It possesses
a cohomological interpretation. Namely,
$\phi_\lambda$ is an alternating bihomomorphism on the stabilizer $\cals$. 
For abelian groups such maps are in one-to-one correspondence to
cohomology classes in $H^2(\cals_\lambda,{\rm U}(1))$.
To any such a cohomology class $[\eps]$ there is associated the twisted
group algebra $\complex_\eps\cals_\lambda$; this is a \findim\ associative
algebra with a basis $b_J$ labelled by $\cals_\lambda$ and multiplication rule
  $$ b_{J_1} \,{\times}\, b_{J_2} = \eps(J_1,J_2)\, b_{J_1 J_2} \, . $$
If the cocycle $\eps$ is cohomologically non-trivial, which is the case if and 
only if $\phi$ is not identically $1$,   i.e.\ if and only if the central
stabilizer $\calc_\lambda$ is strictly smaller than the stabilizer 
$\cals_\lambda$, then the twisted group algebra $\complex_\eps\cals_\lambda$ 
is {\em non-abelian\/}.
It is a direct sum of $|\calc_\lambda|$-many full matrix algebras of rank
  $$ n := \sqrt{ |\cals_\lambda| \,/\, |\calc_\lambda|} \, . $$
(By the cohomology theory of finite abelian groups,
$n$ is an integer.) The fact that the twisted group algebra can be
non-abelian will have important consequences for the gauge symmetries on D-branes.

The boundary labels are now defined as equivalence classes $[\rho,\psi]$
of pairs, where $\rho$ is any label of ${\mathfrak A}$ and $\psi$ is a character
of the central stabilizer. The equivalence relation is the same as for the
labels of simple current extensions (cf.\ Section \ref{scext}), but with
$F_\rho$ replaced by $\phi_\rho$. Accordingly, for boundary conditions
the phenomenon of fixed point resolution arises, too.

To each boundary condition $[\rho,\psi]$ for a simple current invariant of
extension type one associates its {\em automorphism type\/} $g_\rho$. This 
is the character of $\calg'$ that is furnished by the monodromy charge:
  $$ g_\rho(J) = \eE^{2\pi\ii Q_J(\rho)} $$
for $J\iN\calg'$. Recall that the monodromy charge is preserved in fusion 
products of ${\mathfrak A}$. In the annulus coefficients it is preserved as well: 
Expanding the annulus amplitude $A_{[\rho_1,\psi_1],[\rho_2,\psi_2]}$ in 
terms of characters of ${\mathfrak A}$, only characters of monodromy charge 
$Q(\rho_1)\,{-}\,Q(\rho_2)$ contribute.

We are now finally in a position to determine the boundary coefficients
$B_{(\lambda,J),[\rho,\psi]}$ that according to
  \be  |[\rho,\psi]\rangle\!\rangle = \sum_{(\lambda,J)}
  B_{(\lambda,J),[\rho,\psi]}\, |(\lambda,J)\rangle \ee
express the boundary states through the Ishibashi states.
The formula is of amazing simplicity:
  \be  B_{(\lambda,J),[\rho,\psi]} =
  \sqrt{\frac{|\calg'|}{|\cals_\rho|\, |\calc_\rho|}}\,
  \frac{S^J_{\lambda,\rho}}{\sqrt{S_{\Omega,\lambda}}} \, \psi(J)^* \, .
  \labl{foe}

Notice that the boundary coefficients are expressed in terms of chiral
data, in particular the matrices $S^J$ that describe the modular transformations
of the one-point blocks on the torus. The formula \erf{foe} encompasses
several formulae that were previously known in more special
situations; a similar formula can also be derived for the crosscap
\cite{fhssw}. One can show that the matrix $B$ is unitary; this implies
in particular that the number of boundary conditions equals the number of
Ishibashi states. Finally, we mention that from the result \erf{foe} one can
compute the annulus coefficients and show them to be non-negative integers.

\section{Applications to string compactifications}

A major role of full CFT is to provide the world sheet theory of strings.
We now apply the techniques we have just developed to D-branes
in string compactifications. To this end we first study chiral
aspects of string compactifications.

\subsection{Chiral aspects of string compactifications}

We consider compactifications of the string to $D$ flat space-time dimensions.
For simplicity we take $D$ to be even, which includes the most interesting case
$D\eq4$. String theory requires a time-like coordinate on the target space; for
the bosonic string and the usual superstring, one needs Minkowskian signature.
We will therefore require one of the flat dimensions to be time-like.

At the level of chiral CFT the flat dimensions are described by the tensor
product of $D$ free bosons and $D$ free fermions. As we have seen, this
tensor product $\calc_D^{{\rm st,bos}}{\times}\, \calc_{D/2}^{{\rm st,ferm}}$
has $N\eq2$ supersymmetry (subscripts stand for the Virasoro central charge).
The compactification manifold gives rise to another $N\eq2$ superconformal
theory $\calc^{{\rm inner}}$.

Remember that the main idea of (perturbative) string theory is to `forget'
the world sheet and that as part of this task we must take the cohomology
with respect to an $N\eq1$ super-Virasoro algebra. 
We therefore must include in the tensor product also two of the first order
systems that have been discussed in Section 3.3: The chiral CFT 
$\calc^{{\rm gh}}_{-26}$ of ghosts to gauge the stress-energy tensor, and 
the chiral CFT $\calc^{{\rm sgh}}_{11}$ of superghosts to gauge one $N\eq1$ 
supercurrent. We are thus lead to consider the tensor product
  \be  \calc_D^{{\rm st,bos}}\times \calc_{D/2}^{{\rm st,ferm}}
  \times \calc^{{\rm inner}} \times \calc^{{\rm gh}}_{-26}
  \times \calc^{{\rm sgh}}_{11} \, . \labl*

A cohomology can only be defined if the BRST operator constructed from
symmetry generators and ghosts is nilpotent, or in other words, only if the 
chiral symmetry in \erf* is a topological VOA. This requires that the total
Virasoro central charge of the tensor product vanishes. As a consequence,
the central charge of the inner sector is fixed to 
  $$  c = 15\,{-}\,3D/2 \, $$

The tensor product theory can be simplified as follows. We first observe that
neither $\calc_D^{{\rm st,bos}}$ nor $\calc^{{\rm gh}}_{-26}$ contain
space-time fermions; these factors are therefore easy to understand, so
we drop them from our discussion. Next,
we notice that the superghosts can be bosonized, giving rise to a free
boson with `wrong' sign in the two-point blocks. Also, bosonization of the
fermions gives free bosons compactified on the root lattice $L_{D/2}$
of the \lie\ $\mathfrak{so}(D)$. Hence superghosts and space-time fermions
can simultaneously be described by free bosons compactified on a lattice
$L_{D/2,1}$ of Lorentzian signature. For all aspects concerning representations
of the mapping class group, in particular for the modular transformations
of the characters, this chiral non-unitary CFT behaves exactly like the unitary 
theory of $\mathfrak{so}(D{+}6)$. (To get all scalar factors right, one 
should also tensor with $E_8$ at level 1, a subtlety that we will neglect.)
In the correspondence between $L_{D/2,1}$ and $\mathfrak{so}(D{+}6)$,
scalar and vector representations should be exchanged and the spinorial
representations acquire additional minus signs. This mapping goes under the name
{\em bosonic string map\/}. It also plays in important role in the construction
of the heterotic string; for a review see \cite{Lesw}.

In conclusion, we can restrict our attention to the tensor product
  \be \mathfrak{so}(D{+}6) \times \calc^{{\rm inner}} \, . \labl{tens}
It is, however, {\em not\/} simply this tensor product that describes
the chiral aspects of a string compactification. We already saw
that the alignment of the world sheet fermions forces us to perform
a simple current extension; here the relevant simple current is 
$({\rm v},{\rm v})$ where the first field ${\rm v}$ is the primary of 
$\mathfrak{so}(D{+}6)$ that corresponds to the vector
conjugacy class and the second ${\rm v}$ is the primary field that contains
the supercurrent of the inner sector.

Moreover, this is not the only projection we must impose. Space-time
supersymmetry requires the GSO projection.
After the bosonic string map, the GSO projection corresponds to a simple
current extension as well, this time one that is connected to spectral flow.
Indeed, the generators for space-time supersymmetry can be constructed
from the spectral flow operators. Even a converse of this
statement holds \cite{bdfm}: For $D\eq4$, space-time supersymmetry requires
$N\eq2$ superconformal symmetry on the world sheet. (But the situation is
more involved for $D\eq3$ or $2$ \cite{shVa} which geometrically
corresponds to the existence of manifolds with special holonomy.)
The extending simple current can be given explicitly: It is the field
$(s,s)$, where the first $s$ refers to the spinor representation
of $\mathfrak{so}(D{+}6)$ and the other $s$ is the spectral flow simple
current of $\calc^{{\rm inner}}$.
This simple current has integral conformal weight:
  $$    \frac{D+6}{16} + \frac c{24} = \frac{D+6}{16} + \frac{15-3D/2}{24}
  = 1\,.  $$
Here we inserted the value $c$ of the Virasoro central charge for the inner
sector; this value is determined by the nilpotency of the BRST operator.
It is remarkable that this condition (which does not know anything about
space-time supersymmetry) also ensures that we can consistently implement the
GSO projection with simple currents.

Let us summarize our discussion. The chiral CFT underlying a string compactification
can be formulated as the tensor product \erf{tens} {\em together\/} with
a specific simple current extension. This point is already important for
closed string theory, e.g.\ for the correct computation of massless spectra.
It becomes even more crucial once one considers boundary conditions.
Indeed, we learned that chiral data like S-matrix
elements enter crucially in the construction of boundary states; these
chiral data should be the ones of the simple current extension, not the ones of
the naive tensor product \erf{tens}.

\subsection{The Gepner construction}

We are now ready to examine a specific construction of string vacua that 
dates back to Gepner.  His idea was to use a tensor product
  \be  \bigtimes_{\alpha=1}^{r_{\phantom|}}\, \calc^{(\alpha)}  \labl{inner}
of $N\eq2$ minimal models as the starting point for describing the inner 
sector. The levels $k_\alpha$ of the minimal models must be chosen such that
the correct Virasoro anomaly results:
  $$ \sum_{\alpha=1}^r \frac{3 k_\alpha}{k_\alpha+2} = 15-\frac{3D}2 \, . $$
A famous solution for $D\eq4$ is given by $r\eq5$ and all levels equal to 3.
As it turns out, this model (with charge conjugation modular invariant)
corresponds geometrically to a compactification on the so-called Quintic 
threefold, i.e.\ the zero-locus of the homogeneous polynomial $\sum_{\alpha=1}
^5\! X_\alpha^5\eq0$ in the projective space $\complex{\dl P}^4$.

We have seen that one cannot simply take the tensor product \erf{inner} as the 
inner sector, but in addition one must extend by the bilinears in the 
supercurrents so as to obtain a supersymmetric theory. Furthermore, the GSO 
projection is (after the bosonic string map) a projection to even integral 
U(1) charges. By comparison with the possible values of the U(1) charge in the
flat space-time and superghost parts of the string theory, it follows that to
obtain a CFT on which the GSO projection can be performed, we already need to 
project the tensor product \erf{inner} onto integral U(1) charge. (One might 
refer to this operation as the pre-GSO projection.) Thus, the internal theory
used as $\calc^{\rm inner}$ in the sense of the previous subsection really 
is an extension of \erf{inner} by the $(\zet_2)^{r-1}$ group of bilinears
in supercurrents ${\rm v}_i {\rm v}_j$ and by the additional current 
  $$ {\rm v}_1^{D/2-1} (s^2,s^2,...\,,s^2) $$
with $s\eq\Phi^{0,1}_1$ as described at the end of Subsection 4.3.
Only after this extension can the inner theory be compared with the 
$\sigma$-model on a Calabi\hy Yau manifold. Accordingly we call this extension 
of the tensor product the {\em Calabi\hy Yau extension\/}.

When studying string compactifications in the geometrical framework,
one finds that often several {\em different\/} compactification manifolds 
give physically indistinguishable string theories.
In particular, given a string theory compactified
on a CY manifold $\cal M$, there is another CY manifold $\cal W$, the
{\em mirror manifold\/}, that gives the same physical results.

In the conformal field theory description, the transition from $\cal M$ to
$\cal W$ amounts to replacing a modular invariant $Z_{\lambda,\mu}$ by 
the invariant $Z_{\lambda^*_{\phantom i}\!,\mu}$, 
where in the Neveu\hy Schwarz sector the star stands for charge conjugation,
i.e.\ $\lambda^*\eq\lambda^+$, while in the Ramond sector it slightly
differs from conjugation. In the Gepner construction, 
this can be afforded by taking the simple current modular invariant for 
a group of simple currents $\calg_{{\rm GP}}$ that beyond
the simple currents already present in the CY extension also
contains all simple currents of the form
  $$  ({\rm v}_1)^\eps_{} \prod_{\alpha=1}^r (\Phi^{0,2}_0)^{\pi_\alpha} $$
such that
  $$ \sum_{\alpha=1}^r \frac{\pi_\alpha}{k_\alpha+2} + \frac{\eps}2
  \in\zet   \, . $$
Here the numbers $\pi_\alpha$ are integers, defined modulo $k_\alpha{+}2$, and
$\eps$ vanishes for $r\,{+}\,c_{{\rm inner}}/3\iN2\zet$ and can take
the values $0$ and $1$ for $r\,{+}\,c_{{\rm inner}}/3\iN2\zet{+}1$. Moreover, an
appropriate discrete torsion matrix $X$ \erf{dt} must be chosen; for
details we refer to \cite{fkllsw}.
This construction goes under the name {\em Greene\hy Plesser construction}.
It has been most successfully applied in the context of Gepner models
(i.e.\ when one starts from the product of minimal models), where it receives 
its well-known geometric interpretation for the mirror construction. 
However, using simple current symmetries to obtain a  mirror model should also 
work for some other cases, such as for certain Kazama\hy Suzuki models.

\subsection{Applications to D-branes}

We now apply the results of Section \ref{bsc} on boundary conditions
in simple current modular invariants to the study of D-branes in string
compactifications. In string theory, the action of the (super-)Virasoro
algebra is gauged. As a consequence one is not allowed to break
superconformal invariance of the underlying chiral CFT. This implies
that the subalgebra of the chiral algebra that is preserved by the
boundary conditions must contain at least an $N\eq1$ subalgebra of the
total $N\eq2$ algebra as well as the bilinears in the vector currents $v$. We 
refer to such boundary conditions as being {\em super-conformally invariant\/}.

{}From the geometric analysis one knows that D-branes in CY compactifications
come in two species. Those of A-{\em type\/} are sensitive to the symplectic
structure of the CY manifold; they correspond to special Lagrangian submanifolds
(together with U(1)-bundle on them).
Those of B-{\em type\/} test the holomorphic structure of the compactification
space; they correspond to holomorphic vector bundles on submanifolds or,
more generally, to coherent sheaves on the CY. Mirror symmetry exchanges
D-branes of type A on $\cal M$ with D-branes of type $B$ on its mirror $\cal W$.
(This fact can be used to give the conceptually clearest definition of
the mirror $\cal W$ as the moduli space of so-called supersymmetric 3-cycles
of $\cal M$; for a review see \cite{morr5}.)

The algebraic characterization uses the U(1) current $J(z)$ of the total $N\eq2$
algebra. For the charge conjugation modular invariant,
D-branes of type A correspond to (super-)conformally invariant
boundary conditions that preserve $J$, i.e.\ we have a
``Neumann-like'' condition on $J$ which translates into
  \be \left( J_n\oT \bfe + \bfe \oT J_{-n}\right ) {|B \rangle}_{\!\rm A} = 0 
  \labl{A}
for the Ishibashi state.
Similarly, for B-type one has ``Dirichlet-like'' boundary conditions:
  \be \left( J_n\oT \bfe - \bfe \oT J_{-n}\right ) {|B \rangle}_{\rm B} = 0 \, .
  \labl{B}

Let us briefly discuss boundary conditions of A-type \cite{fusw}.
As we want to preserve world sheet supersymmetry, we
must consider boundary conditions that preserve much more symmetry,
namely the simple current extension ${\mathfrak A}_{{\rm WS}}$ of the tensor product
\erf{tens} by all bilinears of supercurrents $v$.
(The full CY extension is obtained from ${\mathfrak A}_{{\rm WS}}$ as a further
extension by the cyclic group generated by the spectral flow simple current.)

The following two features of A-type boundary conditions are immediate
consequences of our general discussion in Section \ref{bsc}:
\nxt A simple current orbit $[\rho]$ of ${\mathfrak A}_{{\rm WS}}$-primaries 
     does not suffice to specify an irreducible boundary condition. One needs 
     in addition a character of the (central) stabilizer. In other words, the
     boundary condition associated to an orbit is in general not irreducible 
     and can be split.
     \\
     We will encounter a similar phenomenon for B-type boundary conditions
     as well. In that case more refined geometrical tools allow us
     to relate the splitting to the existence of bound states at threshold.
     By mirror symmetry, we expect the same phenomenon to take place also for
     A-type branes.
\nxt Boundary fields come in families, labelled by their automorphism type.\,%
 \footnote{~Conformal boundary conditions without automorphism type are of
 interest, too; they correspond to non-BPS D-branes. 
 In contrast to the situation for free theories \cite{sen18,Leru}, so far only
 little is known about such boundary conditions for Gepner models.}
     This quantum number provides a grading on the annuli. It follows that only
     such boundary fields $\Psi^{\rho,\rho'}_\lambda$ exist for which
     the automorphism types are related by $g_\lambda\eq g_\rho/g_{\rho'}$.

In theories of closed strings, one must choose a modular invariant torus
partition function to fix the field content of the theory. This amounts
to prescribing a multiplicity $Z_{\lambda,\mu}\iN\zet_{\geq0}$ for each 
possible bulk field $\Phi_{\lambda\mu}$. For boundary fields in theories 
that contain open strings, a similar choice must be made. However, this task
arises {\em not\/} yet at the conceptual level of full CFT, but only once one 
specifies the string vacuum: For each boundary condition $a$ one must choose a
multiplicity $N_a\iN\zet_{\geq0}$, called the {\em Chan\hy Paton 
multiplicity\/} of $a$. Geometrically, this assignment amounts to choosing a 
{\em brane configuration\/}. At the same time, it might be necessary
to include also unoriented strings; these world sheets effectively
serve to (anti-)symmetrize the closed and open string spectrum. 
Insisting that this can be done in a consistent manner results in
a number of consistency constraints on the conformal field theory, like
relations between amplitudes for different surfaces of Euler characteristic 
zero; for details, we refer to section 4 of \cite{sche11}. The assignment of 
Chan\hy Paton multiplicities must obey additional consistency constraints,
which go under the name of `tadpole cancellation'; they ensure in particular
the absence of anomalies in the low energy effective action. In fact,
anomaly cancellation in these models requires a non-trivial extension of
the Green\hy Schwarz formalism; for a review see \cite{sagn2}.

Our previous results already constrain the possible brane configurations.
For instance, for a brane configuration with branes of different
automorphism type $g_\rho\,{\ne}\,g_{\rho'}$, boundary fields with
$g_\lambda\,{\ne}\,1$ appear. However, in the present context the monodromy
charge is the one of the spectral flow simple current and hence
projecting to $g_\lambda\eq1$ is equivalent to the pre-GSO projection. We
conclude that such brane configurations do not respect the pre-GSO
projection in the open string sector. As a consequence, typically tachyonic
open string modes appear and the whole configuration is unstable. This
effect generalizes the behavior of branes at angles in a flat background.

\subsection{D-branes of B-type}

The fact that mirror symmetry exchanges A- and B-type branes allows us
to discuss D-branes of B-type in Gepner models
by studying D-branes of A-type on the mirror model. The latter can be
obtained explicitly from the Greene\hy Plesser construction.

The analysis shows that in a tensor product of $r$ minimal models, the 
boundary conditions $\rho$ of B-type are (partially) labelled by a 
collection of integers $(L_1,L_2,...\,,L_r)$ such
that $0\,{\le}\,L_\alpha\,{\le}\,[k_\alpha/2]$. (A further integer
$M$, related to the preserved space-time supersymmetry, is needed as well;
we will suppress this in the discussion below.)
But this labelling does not yet account for the fixed points under the 
Gree\-ne\hy Ples\-ser simple current group. Resolving the fixed points
correctly, we will be able to give a geometric interpretation both for 
fixed point resolution itself and for the role of the central stabilizer.

Starting from the boundary states with label $\rho$ before fixed point
resolution, one can compute the number $\nu$
of vacuum states in the open string sector; the result is
  \be \nu = A_{\rho,\rho^+}^\Omega =
  \sum_{J\in\calg_{{\rm GP}}} \caln_{\rho\rho^+}^{\,\ \ J} = |\cals_{\rho}| \,.
  \labl{nu}
Here $\caln$ are fusion rules of the tensor product theory and
$\cals$ denotes the stabilizer in the Greene\hy Plesser group $\calg_{{\rm GP}}$.
Concretely, denoting by $\ell$ the number of components $\alpha$ such that
$L_\alpha\eq k_\alpha/2$, one finds $\nu\eq 2^{\tilde \ell}$, where 
$\tilde\ell\eq\ell$ if $n{+}r$ is even and $\tilde\ell\eq{\max}(\ell{-}1,0)$
if $n{+}r$ is odd.

The quantity $\nu$ is of direct physical interest because it counts the number
of gauge bosons on the brane $\rho$. To learn more about these gauge bosons,
we observe that the action of $\cals_\rho$ on the space whose dimension as
given by \erf{nu} is typically only {\em projective\/}. We already learned how 
to control the projectivity of such an action by using the central stabilizer. 
Using also the structure of the corresponding twisted group algebra, we find 
that the number $\tilde\nu$ of U(1) factors in the gauge group is equal to the
number of elements in the central stabilizer,
  $$ \tilde\nu = |\calc_\rho |\,. $$
The U(1) factors correspond to center-of-mass degrees of freedom of the brane,
and therefore $\tilde\nu$ counts the number of constituent branes.
Concretely, we find that $\calu\cong\zet_2$ when either $n{+}r$ is even
and $\ell\eq2,4,...$ or $n{+}r$ is odd and $\ell\eq1,3,...\,$.
In all other cases the central stabilizer is trivial.

It follows that the underlying twisted algebra is a direct sum of
$\tilde \nu$ ideals, each of which is a full matrix algebra of rank $N$, where
  \be \nu= N^2 \tilde \nu \,.  \labl{split}
We have thus found $\tilde\nu$ degenerate branes with ${\rm SU}(N)$ gauge
symmetry. The central stabilizer therefore provides a new mechanism to produce
non-abelian gauge symmetries in type II compactifications.

Equation \erf{split} is strongly reminiscent of similar relations for
branes in orbifolds with discrete torsion. Also,
the number $N$ describes an $N$-fold wrapping of the brane. A
world sheet analysis of global anomalies has shown that such a multiple
wrapping can also be the consequence of a flat (but non-trivial) $B$-field
on a torsion two-cycle of the CY of order $N$. It is thus tempting to
speculate that central stabilizers give a hint on the existence of
torsion cycles in the CY.

One can also compute the Ramond-Ramond (RR) charges of the brane $\rho$.
The following results have been obtained in \cite{fkllsw}
for the Fermat point of the K3 surface in the weighted projective space
$\dl{CP}[1{,}1{,}1{,}3](6)$ which
corresponds to a tensor product of three minimal models, all of level
$k\eq4$. Up to normalization, they are given by the one-point correlator of a
massless RR bulk-field $\Phi_{RR}$ on a disc with boundary condition
$\rho$. This information is encoded in the boundary state:
  $$ q_{RR}(\rho) \sim \langle\Phi_{RR}\rangle =
  \langle\Phi_{RR}|\rho\rangle\!\rangle \sim B_{(RR,J),\rho} \, . $$
Comparing the trace ${\rm tr}_{\calh_{\rho_1\rho_2}} (-)^F$ to the
intersection form, one can establish the correspondence to the geometric
charge lattice and the appropriate normalization. One finds that the
RR charge after fixed point resolution is the RR charge of the unresolved,
reducible brane, divided by the number $\tilde\nu$ of components; in short,
RR charge is equally distributed over the resolved fields.

We thus arrive at the following geometrical interpretation.
A first hint comes from the observation that the simple currents with
fixed points typically have non-integral conformal weight. This suggests that
the effect is not related to singularities of the CY manifold (which would also
affect the bulk theory), but rather to singularities of the gauge bundle. 
One therefore expects a relation to an interesting mathematical question, 
namely the compactification of the moduli space of vector bundles on K3.

To corroborate this conjecture, we study a configuration of RR charge
  $$ v=(r, c_1, r+\mbox{$\frac12$}\, c_1^2 \,{-}\, c_2) = (2,0,2\,{-}\,2k) \,, $$
for which we find $\nu\eq\tilde\nu\eq2$. It splits into two branes of
identical charge $(1,0,1{-}k)$. No line bundle can have such a RR charge,
only so-called strictly semi-stable sheaves $\cale$; these are sheaves
$\cale$ which possess a subsheaf $\cale'$ of the same slope,
  $$ \frac{v(\cale)}{{\rm rk}\, \cale} = \frac{v(\cale')}{{\rm rk}\, \cale'}\,. $$
Indeed, such sheaves only appear for values of the RR charges for which
the moduli space is not compact, and this requires that the greatest common
divisor of the RR charges to satisfy ${\rm gcd}(q_0,q_2,q_4)\,{>}\,1$. Thus
we have collinear RR charges, which are the central charges in the
(space-time) supersymmetry algebra. This is characteristic for bound states
at threshold, and establishes our interpretation of fixed point resolution.

\section{Conclusions}

Our conclusions are short. Conformal field theory has many applications.
One exciting application is to string theory: CFT gives insight
into string theory with branes in the strong curvature regime.
In this regime standard geometric tools risk to break down, and the fact
that CFT provides independent information is definitely much welcome.

Another point we would like to emphasize is that the understanding of \twodim\
CFT has made a lot of progress in the past few years. New tools have become 
available which make computations feasible that would have been beyond reach 
some years ago. We therefore expect that the interplay between CFT and 
geometric methods in string theory will continue to be fruitful.

       \vfill

\section*{Acknowledgments}
These notes are based on lectures given by C.S.\ at the E\"otv\"os Summer 
School on {\em Nonperturbative QFT Methods and Their Applications\/}, Budapest,
August 2000, and lectures given at the Institut Henri Poincar\'e, Paris, 
October 2000. They also appear in the ``Notes de cours du Centre \'Emile Borel, 
Institut Henri Poincar\'e, UMS 839 CNRS-UPMC, {\em Supergravity, Superstrings, 
and M-Theory\/}, 18 September 2000 - 9 February 2001". C.S.\ would like to 
thank the organizers of both events for their kind invitation.
\\
We are grateful to many colleagues for numerous discussions over the last
few years. In particular, we would like to thank our collaborators G.\ Felder, 
J.\ Fr\"ohlich, L.R.\ Huiszoon, P.\ Kaste, W.\ Lerche, C.A.\ L\"utken, 
B.\ Pedrini, and A.N.\ Schellekens for having shared their insight with us.
We also thank P.\ Bordalo, G.\ H\"ohn, P.\ Kaste and I.\ Runkel
for helpful comments on earlier versions of these notes.  
       \newpage

 \newcommand\J[5]   {{\sl #5}, {#1} {#2} ({#3}) {#4} }
 \newcommand\JJ[5]  {{\sl #5} {#1} {#2} ({#3}) {#4} }
 \newcommand\JX[5]  {{\sl #5}, {#1} {#2} ({#3})}
 \newcommand\Prep[2]{{\sl #2}, preprint {#1}}
 \newcommand\PRep[2]  {{\em #2}, {#1}}
 \newcommand\inBO[7]{{\sl #7}, in:\ {\em #1} ({#3}, {#4} {#5}), p.\ {#6}}
 \newcommand\iNBO[4]{{\sl #4}, in:\ {\em #1} ({#2}), p.\ {#3}}
 \newcommand\wb{\,\linebreak[0]} \def\wB {$\,$\wb}
 \newcommand\Bi[1]  {\bibitem{#1}}
 \newcommand\BOOK[4]{{\em #1\/} ({#2}, {#3} {#4})}
 \def\jf    {J.\ Fuchs}
 \def\bams  {Bull.\wb Amer.\wb Math.\wb Soc.}
 \def\comp  {Com\-mun.\wb Math.\wb Phys.}
 \def\foph  {Fortschr.\wb Phys.}
 \def\ijmp  {Int.\wb J.\wb Mod.\wb Phys.\ A}
 \def\jams  {J.\wb Amer.\wb Math.\wb Soc.}
 \def\jgap  {J.\wb Geom.\wB and\wB Phys.}
 \def\jopa  {J.\wb Phys.\ A}
 \def\maan  {Math.\wb Annal.}
 \def\npbp  {Nucl.\wb Phys.\ B (Proc.\wb Suppl.)}
 \def\nuci  {Nuovo\wB Cim.}
 \def\nupb  {Nucl.\wb Phys.\ B}
 \def\phlb  {Phys.\wb Lett.\ B}
 \def\phrl  {Phys.\wb Rev.\wb Lett.}
 \def\phrp  {Phys.\wb Rep.}
 \def\sebo  {S\'emi\-naire\wB Bour\-baki}
 \def\sema  {Selecta\wB Mathematica}
 \newcommand\fscp[2] {\inBO{Fields, Strings, and Critical Phenomena} {E.\
     Br\'ezin and J.\ Zinn-Justin, eds.} \NH{Amsterdam}{1989} {{#1}}{{#2}}}
 \def\AMS    {{American Mathematical Society}}
 \def\Ca     {{Cambridge}}
 \def\CUP    {{Cambridge University Press}}
 \def\NH     {{North Holland}}
 \def\NY     {{New York}}
 \def\OUP    {{Oxford University Press}}
 \def\PL     {{Plenum Press}}
 \def\PR     {{Providence}}
 \def\SV     {{Sprin\-ger Ver\-lag}}
\def\Cft           {Conformal field theory}
\def\class         {classification}
\def\cocon         {coset construction}
\def\Con           {Conformal }
\def\dyd           {Dynkin diagram}
\def\inv           {invariance}
\def\modinv        {modular invarian}
\def\Modinv        {Modular invarian}
\def\q             {quantum }
\def\RI            {Riemann}
\def\stc           {statistic}
\def\stt           {string theory}
\def\sym           {symmetry}
\def\syms          {sym\-me\-tries}
\def\voa           {vertex operator algebra}

\end{document}